\documentclass[conference]{IEEEtran}

\usepackage[table]{xcolor}
\usepackage{amsmath,amssymb}
\usepackage{cite}
\usepackage{graphicx}
\usepackage{booktabs}
\usepackage{multirow}
\usepackage{pifont}
\usepackage{enumitem}
\usepackage{xspace}
\usepackage{listings}
\usepackage{balance}
\usepackage[most]{tcolorbox}
\usepackage{array}
\usepackage{tikz}
\usepackage{float}
\usepackage{url}
\usepackage[hidelinks]{hyperref}

\definecolor{HeaderBlueGray}{HTML}{7EA4B6}
\definecolor{HeaderGreen}{HTML}{71A682}

\newcommand{\CircleHeader}[2]{%
  \tikz[baseline=(char.base)]{%
    \node[
      shape=circle,
      fill=#1,
      text=white,
      font=\bfseries\small,
      inner sep=0pt,
      minimum size=0.95em
    ] (char) {#2};%
  }%
}
\newcommand{\HeaderBlack}[1]{\CircleHeader{black}{#1}}
\newcommand{\HeaderBlueGray}[1]{\CircleHeader{HeaderBlueGray}{#1}}
\newcommand{\HeaderGreen}[1]{\CircleHeader{HeaderGreen}{#1}}

\newtcolorbox{summarybox}{
  colback=gray!10, frame hidden, boxrule=0pt,
  arc=1mm, left=4pt, right=4pt, top=2pt, bottom=2pt,
  width=\linewidth,
}

\newcommand{\sysname}{\textsc{SIGIL}\xspace}
\newcommand{\eg}{e.g.,\xspace}

\newcommand{\etal}{et~al.\xspace}

\setlist[enumerate,itemize,description]{leftmargin=*}

\providecommand{\Description}[1]{}

\begin{document}

\title{Sealing the Audit--Runtime Gap for LLM Skills}


\author{%
\IEEEauthorblockN{%
Tingda Shen\IEEEauthorrefmark{1},
Yebo Feng\IEEEauthorrefmark{2},
Konglin Zhu\IEEEauthorrefmark{1},
Xiaojun Jia\IEEEauthorrefmark{2},
Yang Liu\IEEEauthorrefmark{2}, and
Lin Zhang\IEEEauthorrefmark{1}}
\IEEEauthorblockA{\IEEEauthorrefmark{1}%
Beijing University of Posts and Telecommunications, Beijing, China\\
\{shentingda, klzhu, zhanglin\}@bupt.edu.cn}
\IEEEauthorblockA{\IEEEauthorrefmark{2}%
Nanyang Technological University, Singapore\\
\{yebo.feng, yangliu\}@ntu.edu.sg, jiaxiaojunqaq@gmail.com}}

\IEEEoverridecommandlockouts
\makeatletter
\def\@IEEEpubidpullup{6.5\baselineskip}
\makeatother
\IEEEpubid{\parbox{\columnwidth}{%
		Network and Distributed System Security (NDSS) Symposium 2027\\
		22--26 March 2027, Seoul, Republic of Korea\\
		ISBN 978-1-970672-09-1\\
		https://dx.doi.org/10.14722/ndss.2027.[23$|$24]xxxx\\
		www.ndss-symposium.org
}
\hspace{\columnsep}\makebox[\columnwidth]{}}

\maketitle

\begin{abstract}
Large language model (LLM) ecosystems such as Claude Code and ChatGPT
increasingly rely on skills: packages of natural-language instructions
and executable tools. Once in the LLM's context, skill content cannot
be reliably separated from trusted instructions, and a skill's
executable side can invoke privileged actions, exposing the skill
supply chain to injection, tampering, and rug-pull attacks. Existing
defenses are stage-bound: centralized signing, audit reports unbound
from the runtime artifact, or policy engines that cannot attest to what
was approved.
We present \sysname, the first framework that seals the audit--runtime
gap for LLM skills. \sysname delivers \emph{verifiable hosting} through
a backend-agnostic, tamper-evident registry from which LLMs fetch
skills directly. The registry admits four publication types,
Transparent, Licensed, Sealed, and Committed, spanning plaintext public
distribution, monetized access, custodial use, and internal workflows; before admission, every
skill is vetted by a Decentralized Autonomous Organization (DAO) audit
committee that supports pluggable auditing methods under a
stake-and-slash economic model. At load time, \sysname delivers
\emph{verified loading} through a skill verification protocol executed
by a Skill Verification Loader (SVL) embedded as the mandatory loading
path: the SVL retrieves and decrypts the skill as its type requires,
verifies its integrity against the tamper-evident registry record, and
enforces its permission manifest before context injection. We evaluate
\sysname in a real-world deployment against six representative attack classes using
1{,}079 in-the-wild skills, which we collect and release as the open benchmark for end-to-end skill defense. We instantiate its registry over Ethereum
Sepolia, immudb, Tessera, and Solana. At publication time, \sysname achieves
strong detection accuracy. At load time, the SVL verifies each skill's
integrity against its registry record and enforces its approved
permission manifest, while registry invocation latency ranges from 51.7 to 157.6\,ms across the four backends.
 Together, these results show that LLM skills can be
cryptographically bound from publication through runtime at practical
cost.
\end{abstract}


\IEEEpeerreviewmaketitle

\section{Introduction}
\label{sec:intro}

\begin{figure}[!t]
  \centering
  \includegraphics[width=0.95\linewidth]{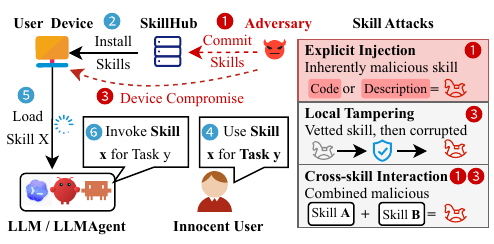}
  \caption{Three skill-level attack categories: explicit injection, local tampering, and cross-skill interactions. The full
  six-attack taxonomy is in \S\ref{sec:threat-model}.}
  \Description{Three skill-level attack categories are summarized: explicit injection at submission time, local tampering after approval, and cross-skill interaction during execution.}
  \label{fig:skillattack}
\end{figure}

The rise of large language models (LLMs) has introduced \emph{skills}: packages of natural-language instructions and executable tools
that allow agents to invoke APIs, manipulate file systems, and interact with other real-world services~\cite{iqbal2023llmplatform}.
Access to skills relies on official and third-party platforms, such as the GPT Store, Claude Marketplace, and ClawHub.
Holzbauer~\etal~\cite{holzbauer2026malicious} collected 238{,}180 unique skills from three major distribution platforms and GitHub,
providing empirical evidence of a large-scale ecosystem in which thousands of developers publish skills that agents can dynamically discover and invoke.

Alarmingly, the skill ecosystem is already exposed to supply-chain attacks at the marketplace level.
Jiang~\etal~\cite{jiang2026harmfulskillbench} identified more than 4{,}858 malicious skills on skill distribution platforms,
while a scan of 3{,}984 skills from ClawHub found that 36.82\% of skills contained at least one security issue and identified 1{,}467 malicious payloads~\cite{snyk2026toxicskills}.
The severity of this threat stems from the privileged role skills play in agent execution: once installed or retrieved, a skill can inject persistent
instructions, steer tool selection, execute code, and access user-side resources under the agent's trust assumptions.
These findings reveal a systemic skill supply-chain threat rooted in the way LLM agents discover, trust, and execute third-party skills.

The mechanism of skill-level attacks is that, once in the LLM's context, a skill's natural-language description and schema cannot be reliably separated from trusted instructions,
and its executable side can invoke privileged actions through APIs, databases, file systems, and other real-world services.
As a result, a carefully crafted malicious description can subvert the agent's behavior without altering any executable code.
The attack surface therefore extends beyond traditional code-level vulnerabilities to encompass the descriptions and schemas
that both influence agent reasoning and drive tool invocation, through which agents select, load, and use skills.
Skill attacks can be launched at multiple stages of a skill's
lifecycle, including distribution, publication, storage, and
loading, giving current systems a long vulnerability surface. Depending on the
stage at which they occur, we distinguish six categories: \emph{explicit injection}, \emph{implicit poisoning}, \emph{rug pull}, \emph{cross-skill interaction}, \emph{auditor collusion}, and \emph{local tampering}; we discuss them in detail in \S\ref{sec:threat-model}. Figure~\ref{fig:skillattack} illustrates three of the six categories.

Existing defenses for skills can be organized into four technical categories across three lifecycle stages.
At the submission stage, static scanning methods inspect skill files and bundled code for known risky patterns, such as suspicious commands, secret access, or unsafe network operations~\cite{liu2026agentskillswild}.
Semantic detection methods analyze the intended behavior of skills from natural-language instructions and heterogeneous artifacts, using LLM-based classification, behavioral verification, or graph reasoning to detect prompt injection and malicious workflows~\cite{liu2026maliciousskillswild,wang2026malskills,li2025dissonances}.
At the anchoring stage, supply-chain governance methods propose integrity verification~\cite{jamshidi2025securing}, permission models~\cite{zhu2025miniscope}, marketplace vetting~\cite{liu2026agentskillswild,iqbal2023llmplatform,yan2024chatgpt,hou2025security}, and trust-tiered execution to secure skill distribution, yet these mechanisms remain early-stage and are not consistently enforced in current registries~\cite{jiang2026sokskills,bhardwaj2026formal,li2026towards}.
At the invocation stage, runtime protection methods monitor or constrain agent behavior during skill execution~\cite{shi2025progent,zhu2025miniscope}, such as filtering external inputs, checking tool calls, or isolating execution, but they cannot prevent malicious skills from entering marketplaces~\cite{zhao2026clawguard}.
Existing defenses cover different stages of the skill lifecycle, but none provides end-to-end protection from skill auditing and publication, through marketplace provenance, to runtime loading.

To address this gap, we propose \sysname, the first framework to seal the audit--runtime gap for LLM skills, providing end-to-end protection from skill submission to invocation: every submitted skill is vetted before invocation, every stored skill is tamper-evident, every loaded skill is fully verified before execution, and every step of this lifecycle is publicly traceable.
Specifically, \sysname operates in three stages: \emph{Submission}, \emph{Anchoring}, and \emph{Invocation}. \sysname instantiates these stages on a backend-agnostic registry that provides tamper-evident storage and public traceability. The registry supports immutable-database, blockchain,
and transparency-log backends without altering the protocol.
At \emph{Submission}, a developer commits a skill to a Skill Pre-Registry, after which a Decentralized Autonomous Organization (DAO)~\cite{jungnickel2024dao} audit committee reviews it through reputation-weighted consensus; the committee is pluggable, allowing auditors to apply LLM-based vetting~\cite{jamshidi2025securing}, sandbox testing~\cite{debenedetti2024agentdojo}, or any combination, while a \emph{stake-and-slash} mechanism penalizes non-consensus or retrospectively overturned votes.
At \emph{Anchoring}, approved skills migrate from the Pre-Registry to a tamper-evident Skill Registry that stores each skill's content (or content hash), permission manifest, audit report, and version history. The registry supports four publication types: \emph{Transparent} (plaintext in the registry), \emph{Licensed} (ciphertext in the registry with paid access), \emph{Sealed} (ciphertext in the registry with developer-held keys), and \emph{Committed} (hash-only in the registry, content retained locally). Together, these publication types support diverse distribution and access requirements while protecting developers' intellectual property and enabling the sharing of high-quality, commercial-grade skills.
At \emph{Invocation}, the Skill Verification Loader (SVL), integrated into the AI Service Provider, acts as the sole entry point for skill loading: it queries the registry, hashes content as needed, and cryptographically binds the loaded skill to its audited record, complementing existing runtime defenses such as Progent~\cite{shi2025progent} and MiniScope~\cite{zhu2025miniscope}.

We evaluate \sysname on 1{,}079 in-the-wild skills against six external defenses. \sysname is the only defense that provides coverage across all six representative attacks: 97.6\% accuracy on explicit injection and implicit poisoning, 95.1\% on rug pull, 100\% on local tampering, 90.2\% mitigation on cross-skill interaction, and stable accuracy under 20--40\% colluding auditors; the DAO Committee reaches 97.6\% accuracy and 1.5\% false-negative (FN) rate. In terms of overhead, mean invocation latency ranges from 51.7 to 157.6 ms across the four registry backends, and audit token usage stays under 3\% of a typical \$20 monthly LLM subscription. A game-theoretic analysis and a 600-round simulation show that honest auditing is the unique Nash equilibrium and the dominant long-run strategy.

This paper makes the following contributions:
\begin{enumerate}
    \item \textbf{An end-to-end lifecycle security framework for LLM skills.} We present \sysname, the first skill-security framework that protects the integrity and provenance of LLM skills across their full lifecycle, sealing the audit--runtime gap so that what is audited is exactly what the agent later loads.
    \item \textbf{Multi-type publication mechanism for skill IP protection.} We introduce multiple skill publication types that differ in payment and cryptographic protection, thereby protecting developers' intellectual property.
    \item \textbf{A cryptographically verified runtime loader.} We design the Skill Verification Loader (SVL), a single agent-side integration point that hides registry and publication-type complexity, binds each loaded skill to its audited record, blocks post-audit tampering, and adds at most 0.495\,ms of local verification overhead.
    \item \textbf{An economic model for sustainable operations.} A token-credit settlement aligned with stake-and-slash incentives, designed so that honest participation remains profitable over time and the ecosystem sustains itself without a central operator.
\end{enumerate}


\section{Related Work}
\label{sec:related}

\subsection{Attack Vectors on LLM Skills}
\label{sec:bg-attacks}

LLM skills face risks at every lifecycle stage~\cite{li2026towards}
and inherit supply-chain risks from conventional software
ecosystems~\cite{wang2025large,gu2023investigating,ladisa2023sok}.
Figure~\ref{fig:skillattack} groups skill attacks into three lifecycle
categories: \emph{Explicit Injection} at \textbf{submission},
\emph{Local Tampering} after \textbf{anchoring}, and
\emph{cross-skill interaction} during \textbf{invocation}.

Malicious skill content embeds adversarial payloads in descriptions
or metadata. Because LLMs cannot separate instructions from
data~\cite{debenedetti2024agentdojo,yi2025benchmarking,yu2025trustworthy}, prompt
injection underlies most skill-level
attacks~\cite{schmotz2026skill}: \emph{direct injection} places
instructions in user input~\cite{10.1145/3773080};
\emph{indirect injection} hides them in external data the agent later
retrieves~\cite{greshake2023indirect,an2025ipiguard}; \emph{tool
poisoning} embeds them in a skill description so the LLM follows them
during tool selection, even when the skill is never
executed~\cite{wang2025unique,wang2025mcptox}; and \emph{metadata
attacks}~\cite{mo2026attractive} optimize metadata to raise the chance of
invocation.

Rug pull attacks silently modify a skill after it passes
review~\cite{jamshidi2025securing}; when the registry stores only a
mutable reference (\eg a URL), such modifications evade re-inspection
while users still trust the approved version.

Cross-skill interaction attacks exploit the shared agent context:
individually benign skills jointly induce harm.
\emph{Shadowing} mimics a legitimate skill's name to hijack
invocations~\cite{song2025beyond}, and the \emph{Puppet Attack}
injects control directives that redirect subsequent tool
calls~\cite{zong2025mcpsafetybench}. Per-skill scanning is therefore
insufficient~\cite{li2026towards}.

\begin{table}[t]
\definecolor{relGray}{gray}{0.94}
\newcommand{\relCheck}{\begingroup\large\ding{51}\endgroup}
\newcommand{\relCheckRed}{\begingroup\large\color{red}\ding{51}\endgroup}
\newcommand{\relCross}{\begingroup\large\ding{55}\endgroup}
\newcommand{\relHeadTwo}[2]{\shortstack[c]{\textbf{#1}\\\textbf{#2}}}
\newcommand{\relHeadThree}[3]{\shortstack[c]{\textbf{#1}\\\textbf{#2}\\\textbf{#3}}}
\newcolumntype{Y}{>{\centering\arraybackslash}m{6.4em}}
\caption{Capability comparison of representative defenses.}
\label{tab:bg-defense-capabilities}
\centering
\normalsize
\setlength{\tabcolsep}{4pt}
\setlength{\arrayrulewidth}{0.4pt}
\renewcommand{\arraystretch}{1.20}
\rowcolors{0}{gray!12}{white}
\resizebox{\columnwidth}{!}{%
\begin{tabular}{l|YYYY}
\toprule
\rowcolor{white}
\textbf{Defense} &
\relHeadThree{Submission}{stage}{screening} &
\relHeadThree{Anchoring}{stage}{integrity} &
\relHeadThree{Invocation}{stage}{enforcement} &
\relHeadTwo{Version}{traceability} \\
\midrule
SkillScan~\cite{liu2026agentskillswild} & \relCheck & \relCross & \relCross & \relCross \\
Semantic auditing~\cite{liu2026maliciousskillswild,wang2026malskills} & \relCheck & \relCross & \relCross & \relCross \\
Chord~\cite{li2025dissonances} & \relCheck & \relCross & \relCross & \relCross \\
Signed MCP packages~\cite{jamshidi2025securing} & \relCheck & \relCheck & \relCheck & \relCross \\
Trust-tiered governance~\cite{jiang2026sokskills,bhardwaj2026formal,li2026towards} & \relCheck & \relCross & \relCross & \relCross \\
Mainstream markets~\cite{liu2026agentskillswild,iqbal2023llmplatform,yan2024chatgpt,hou2025security} & \relCheck & \relCross & \relCross & \relCross \\
Progent~\cite{shi2025progent} & \relCross & \relCross & \relCheck & \relCross \\
MiniScope~\cite{zhu2025miniscope} & \relCross & \relCross & \relCheck & \relCross \\
MindGuard~\cite{wang2025mindguard} & \relCross & \relCross & \relCheck & \relCross \\
ClawGuard~\cite{zhao2026clawguard} & \relCross & \relCross & \relCheck & \relCross \\
AgentSentinel~\cite{hu2025agentsentinel} & \relCross & \relCross & \relCheck & \relCross \\
AgentFuzz~\cite{liu2025make} & \relCross & \relCross & \relCheck & \relCross \\
\textbf{\sysname} & \relCheckRed & \relCheckRed & \relCheckRed & \relCheckRed \\
\bottomrule
\end{tabular}%
}
\end{table}

\subsection{Existing Defenses}
\label{sec:bg-defenses}

We organize prior defenses by the three lifecycle stages of
§\ref{sec:bg-attacks}~\cite{li2026towards}.

Submission-stage defenses inspect skills before admission. Static
methods scan files and bundled code for risky
patterns~\cite{liu2026agentskillswild}; semantic methods analyze
descriptions via LLM classification, behavioral verification, or
graph reasoning~\cite{liu2026maliciousskillswild,wang2026malskills},
and \textsc{Chord}~\cite{li2025dissonances} scans tool definitions
for Cross-Tool Harvesting and Polluting (XTHP) attacks; structured
queries also resist prompt injection~\cite{chen2025struq}. None of
these methods binds the verdict cryptographically to the bytes later
loaded by the agent.

{\sloppy
Anchoring-stage defenses govern skill storage and trust.
Jamshidi~\etal~\cite{jamshidi2025securing} sign Model Context
Protocol (MCP) packages with RSA and add LLM vetting plus runtime
guardrails, but rely on a centralized signer.
Trust-tiered governance \cite{jiang2026sokskills,bhardwaj2026formal,li2026towards}
structures registry screening but remains inconsistently enforced.
Audits of mainstream marketplaces \cite{liu2026agentskillswild,iqbal2023llmplatform,yan2024chatgpt,hou2025security}
report false negatives and undetected post-publication modifications.
None binds publication provenance to a tamper-evident registry record.
\par}

{\sloppy
Invocation-stage defenses constrain agent behavior when a skill is
loaded. Policy-based methods restrict tool calls or assign
least-privilege permissions at
runtime~\cite{shi2025progent,zhu2025miniscope}; dependency-graph
defenses constrain agent execution against tool-invocation
attacks~\cite{an2025ipiguard}; runtime monitors detect abnormal
behavior and intercept dangerous file, process, and operations~\cite{wang2025mindguard,zhao2026clawguard,hu2025agentsentinel};
directed fuzzing drives attacker-controlled prompts toward sensitive
sinks~\cite{liu2025make}; and adversarial agents and graph neural
networks discover and counter
attacks~\cite{he-etal-2025-red,wang2026malskills,wang-etal-2025-g}.
These defenses reduce runtime harm but remain post-admission: they
neither prevent malicious entry nor bind loaded skills to audited
artifacts. Benchmarks and surveys~\cite{ye2024toolsword,zhan2024injecagent,guo2025systematic,zong2025mcpsafetybench}
reinforce the need for full-lifecycle protection.
\par}

Table~\ref{tab:bg-defense-capabilities} summarizes this fragmentation:
prior defenses are stage-bound and largely centralized.
\sysname instead combines a tamper-evident Skill Registry, a DAO audit committee
with stake-and-slash incentives, and an SVL that binds
every loaded skill to its audited record, covering all three stages and
version traceability in one framework.

\section{Background and Problem Statement}
\label{sec:problem}

\subsection{Background}
\label{sec:bg-skills}

LLM skills are packages of natural-language instructions and executable tools that let agents call APIs, access file systems, and interact with external services~\cite{iqbal2023llmplatform}.
Although packaging differs across platforms, what the model consumes is still text: the skill name, description, parameter schema, and usage instructions enter the agent's context and guide tool selection and execution.
The agent must rely on these textual fields to decide when and how to invoke a skill, even when the skill wraps benign code.

Because LLMs cannot reliably separate instructions from data, prompt-level attacks extend to skill-level attacks.
A malicious skill description can steer the agent before any code runs, and poisoned metadata can raise the chance that a harmful skill is selected~\cite{greshake2023indirect,wang2025unique,wang2025mcptox,mo2026attractive}.
The security problem therefore extends beyond executable payloads: prompt injection, tool poisoning, and metadata manipulation all exploit how skills are trusted as text inside the model's context~\cite{schmotz2026skill,li2026towards}.

\subsection{Threat Model}
\label{sec:threat-model}

\begin{table*}[t]
\caption{The four publication types of \sysname skills, grouped by
public distribution and the developer's exclusive use.}
\label{tab:skilltypes}
\centering
\small
\setlength{\tabcolsep}{3pt}
\renewcommand{\arraystretch}{1.15}
\resizebox{\textwidth}{!}{%
\begin{tabular}{@{}m{2.2cm}m{3cm}m{7cm}m{2.2cm}m{1.8cm}m{1.5cm}@{}}
\toprule
\textbf{Type} & \textbf{Usage scope} & \textbf{Description} & \textbf{Registry content} & \textbf{Access control} & \textbf{Load source} \\
\midrule
\includegraphics[height=1.15em]{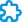}\hspace{0.15em}Transparent
& \multirow{2}{=}[-1ex]{\raggedright\arraybackslash Public distribution}
& Plaintext and free; anyone can inspect and reuse the skill.
& Plaintext
& None
& Registry \\
\cmidrule(r){1-1}\cmidrule(l){3-6}
\includegraphics[height=1.15em]{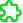}\hspace{0.15em}Licensed
&
& Copyrighted; access granted per paid license.
& Ciphertext
& Purchase
& Registry \\
\midrule[\heavyrulewidth]
\includegraphics[height=1.15em]{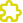}\hspace{0.15em}Sealed
& \multirow{2}{=}[-1ex]{\raggedright\arraybackslash Developer's exclusive use}
& Hosted on the registry; decryptable only by the developer.
& Ciphertext
& Key custody
& Registry \\
\cmidrule(r){1-1}\cmidrule(l){3-6}
\includegraphics[height=1.15em]{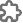}\hspace{0.15em}Committed
&
& Hosted on a local device; only its hash is registered.
& Hash
& Local custody
& Local file \\
\bottomrule
\end{tabular}%
}
\end{table*}

This section defines the threat model for \sysname: the actor classes
and trust boundaries, the attacker's goals and capabilities, and six
representative attacks across the submission, anchoring, and
invocation stages of \sysname.

The \sysname system model has four classes of actors. (1)
\textbf{Skill developers} publish skills to the append-only registry
and may be benign or malicious. (2) \textbf{Auditors} review
submitted skills and stake tokens to vote on approval; individual
auditors may be malicious, but we assume the reputation-weighted
majority remains honest at any given time. (3) \textbf{Users}
discover, install, and invoke skills through their LLM agents and
rely on the trusted audit outcome recorded in the append-only registry rather than re-verifying skill
content. (4) \textbf{LLMs or LLM agents} execute skills on behalf of
users and are treated as untrusted instruction followers that may
obey poisoned descriptions once such content enters their context.
We assume the integrity guarantees of the selected registry backend: accepted records are tamper-evident and cannot be modified without detection.

The adversary aims either to push a malicious skill through DAO
audit into the registry, or to alter an approved skill without
triggering re-audit. To achieve these goals, the
adversary may submit arbitrary skills, register Sybil staked auditor
identities to bias voting, modify a published skill after approval
when it is not immutably anchored in the registry, craft poisoned
descriptions and adversarial metadata to bias the LLM's tool
selection, or compromise an end-user device to tamper with a
locally cached skill so that what the agent loads diverges from the
audited registry record. We assume the adversary cannot violate the
selected backend's integrity or finality guarantees, cannot corrupt
the reputation-weighted majority of the auditing stake, and cannot
replace the LLM's weights or the SVL
integrated into the AI Service Provider.

From the goals and capabilities above, we define six representative
attacks.
(1) \textbf{Explicit injection} embeds overtly harmful instructions
in a skill before publication, so the malicious behavior is present
at submission time and should be detectable by content-based
auditing.
(2) \textbf{Implicit poisoning} hides malicious intent in
benign-looking descriptions or adversarially optimized metadata
that bias the LLM's tool selection~\cite{wang2025mcptox,mo2026attractive}; the
skill looks acceptable, but its wording or metadata raises the
chance that the agent invokes it in unsafe ways.
(3) \textbf{Rug pull} involves passing a benign skill through audit and then
silently performing publisher-side or registry-side modification of the
published artifact after approval, opening a gap between what was audited
and what is later loaded.
(4) \textbf{Cross-skill interaction} combines individually benign
skills so that harm emerges only when they interact through the
shared LLM context~\cite{song2025beyond,zong2025mcpsafetybench}; no
single skill is dangerous in isolation, and the harm comes from
their runtime composition.
(5) \textbf{Auditor collusion} controls or coordinates staked
auditor identities to approve a malicious skill, targeting the
trust decision rather than the skill content; the registry may then
certify a malicious skill as legitimate.
(6) \textbf{Local tampering} modifies a cached or locally stored approved skill through a user-side
tampering path after audit but before
invocation, so the local artifact diverges from the audited registry
record while the registry still treats the skill as
approved.

\subsection{Design Goals}
\label{sec:tm-goals}

\sysname is designed to satisfy four goals.
\textbf{Integrity}: an approved skill version remains immutable, its
content, manifest, and metadata bound to its content-addressed
identifier until a new audit cycle is initiated.
\textbf{Auditability}: each skill version carries an immutable,
publicly verifiable record of reviewer identities, individual votes,
and the final outcome.
\textbf{Accountability}: auditor participation carries explicit
economic responsibility, with each vote subject to staking, reward,
and slashing.
\textbf{Least privilege}: each skill executes with only the minimum
permissions declared at publication time and authorized by the user.

\section{Methodology}
\label{sec:design}

\subsection{Overview}
\label{sec:design:arch}

This section presents the design and workflow of \sysname, whose primary
goal is to ensure that every skill loaded by an LLM agent matches the exact
artifact approved at publication time. Its secondary goal is to integrate
skill security auditing into the admission path, filtering malicious or
non-compliant skills before they are admitted to the Skill Registry.

To accommodate diverse developer and user scenarios, \sysname supports four
publication types, listed in Table~\ref{tab:skilltypes}, that span a
disclosure spectrum from public plaintext to a local-only commitment.
\emph{Transparent} and \emph{Licensed} skills are intended for public
distribution: the former publishes plaintext to the registry for free
inspection and reuse, whereas the latter
publishes ciphertext and gates decryption behind a purchase, letting the
developer retain copyright and earn revenue per use. \emph{Sealed} and
\emph{Committed} skills are reserved for the developer's exclusive use:
the former publishes ciphertext to the registry with decryption keys held
in developer custody, whereas the latter records only a cryptographic
commitment (a content hash, in our implementation) with the registry
and keeps the skill file on the developer's local device. The registry itself is
conceptually a tamper-evident storage abstraction. Our implementation instantiates this abstraction
over multiple heterogeneous backends, including blockchains and append-only databases.

\begin{figure*}[t]
  \centering
  \includegraphics[width=0.96\linewidth]{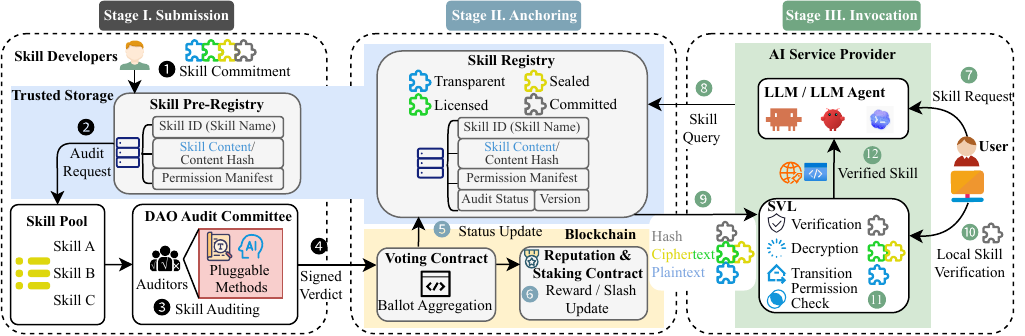}
  \caption{\sysname's three-stage architecture. A developer commits a
    skill (\textbf{Submission}), the DAO Audit Committee vets and anchors
    it (\textbf{Anchoring}), and users load it through the SVL
    (\textbf{Invocation}).}
  \Description{System workflow from skill commitment to the pre-registry, DAO audit and voting, registry anchoring, and SVL-verified loading by users.}
  \label{fig:architecture}
\end{figure*}

\sysname's workflow proceeds in three stages
(illustrated in \figurename~\ref{fig:architecture}), each implemented by a dedicated set
of modules. \textbf{Stage I (Submission)} records the developer's skill
commitment in the Skill Pre-Registry and dispatches the submission to the
DAO Audit Committee, whose auditors analyze the skill independently
through pluggable methods and return signed verdicts.
\textbf{Stage II (Anchoring)} aggregates the verdicts through the Voting
Contract, writes the audit outcome to the Skill Registry, and applies
the corresponding reward or slash to participating auditors via the
Reputation \& Staking Contract. \textbf{Stage III (Invocation)} routes
every user request through the SVL, which
performs type-specific verification and injects the verified skill into the
LLM or LLM agent.

\subsection{Submission Stage}
\label{sec:design:publication}
The submission stage commits a developer's skill to an immutable
reference state, against which the DAO Audit Committee inspects it before
admission. The stage proceeds through four steps: the developer commits
the skill to the Skill Pre-Registry; the registry emits an audit request
that dispatches the submission to the Skill Pool; auditors independently
analyze the skill through pluggable methods; and each auditor returns a
signed verdict.

The registry, audit pool, and verdict log rely on a tamper-evident
storage abstraction. \sysname decouples this abstraction from the
underlying backend, allowing blockchain, append-only database, and
transparency-log implementations without changing the protocol.
In \sysname, the DAO serves as the audit committee. Blockchain remains necessary
for the DAO and economic components, where smart contracts transparently enforce
auditor staking, voting, rewards, and slashing without a central administrator.

The submission process begins with
\textbf{\HeaderBlack{1} Skill Commitment.} The Skill Pre-Registry, the
submission entry point for every submission, records each skill before
admission to the Skill Registry; this binding ties every audit to a
fixed reference. A submitted skill carries the following fields:

\begin{itemize}
\item \textbf{Skill content.} The skill payload, recorded in the form
  prescribed by the publication type: plaintext for \emph{Transparent}
  skills; ciphertext together with the plaintext hash for \emph{Licensed}
  and \emph{Sealed} skills; and a content hash for \emph{Committed}
  skills, whose plaintext file is retained on the developer's local
  device. We refer to this registry payload as $\mathit{content}$ in
  the rest of this section.

\item \textbf{Skill ID.} A content-addressed identifier computed as
\[
\begin{aligned}
\mathit{skill\_id}
= H(&\mathit{content} \parallel \mathit{developer} \parallel \\
   &\mathit{prev\_version} \parallel \mathit{timestamp}),
\end{aligned}
\]
  where $H$ is a collision-resistant hash function. Because
  $\mathit{skill\_id}$ is derived from these fields, any modification
  (notably to $\mathit{content}$) produces a distinct identifier,
  making the registry inherently tamper-evident.

\item \textbf{Permission manifest.} A structured registry declaration
  of the skill's intended capabilities, comprising
  $\mathit{declared\_tools}[\,]$, $\mathit{data\_scope}[\,]$, and
  $\mathit{behavior\_bounds}$. Anchoring the manifest to the tamper-evident record enables
  permission checks during both audit and invocation.
\end{itemize}

Once the commitment is recorded, the protocol issues an
\textbf{\HeaderBlack{2} Audit Request.} The Skill Pre-Registry emits a
\texttt{SkillAudit} event carrying the $\mathit{skill\_id}$ and metadata,
and this event places the submission in the Skill Pool, a queue that
isolates pending skills from approved ones during review. Auditors
claim the task by staking additional tokens as a commitment deposit, and
at least $N$ auditors (configurable; \eg $N=5$) must claim before the
audit proceeds.

Once enough auditors have claimed the task, they perform
\textbf{\HeaderBlack{3} Skill Auditing.} Each auditor independently
analyzes the submitted skill with a method of their choice. The audit
framework is \emph{pluggable} by design: auditors may apply static
analysis, LLM-based semantic review, sandbox testing, or any
combination of these. Decoupling the audit protocol from specific
detection techniques lets \sysname adopt new methods as the field
advances, without changing the protocol itself. To deter dishonest
audits and to encourage auditors to deploy more advanced methods,
\sysname couples the audit framework with an economic model formalized
in \S\ref{sec:design:economics}.

Auditing of \emph{Licensed}, \emph{Sealed}, and \emph{Committed} skills
relies on a direct content-delivery protocol, since plaintext is
never stored in the registry. For \emph{Licensed} and \emph{Sealed} skills,
the registry holds an encrypted blob that each auditor decrypts via
per-auditor key delivery; for \emph{Committed} skills, the registry
holds only the content hash, and the developer transmits the encrypted
skill content to auditors directly through the same key
exchange. In all
three cases, each auditor registers an asymmetric public key
$\mathit{pk}_i$ with the registry at registration time, and the developer
derives a per-auditor delivery key via Elliptic Curve
Diffie--Hellman (ECDH)~\cite{haakegaard2015elliptic,hao2022sok} followed by an HMAC-based Key Derivation
Function (HKDF)~\cite{krawczyk2010cryptographic}:
\begin{align*}
  S_i &= \mathit{sk}_d \cdot \mathit{pk}_i, \\
  k_i &= \mathrm{HKDF}(S_i,\; \texttt{audit},\;
         \mathit{skill\_id} \| \mathit{pk}_i \| \mathit{pk}_d),
\end{align*}
where $\mathit{sk}_d$ is the developer's private key, $\mathit{pk}_d$
is the corresponding developer public key, and $\mathit{pk}_i$ is the
public key of auditor $i$. The developer encrypts the
symmetric content key $k_{\mathit{content}}$ under $k_i$ using AES-GCM
and stores the wrapped key in the registry (for \emph{Licensed} and
\emph{Sealed} skills) or delivers it directly (for \emph{Committed}
skills); each auditor recovers $k_{\mathit{content}}$ with their
private key and decrypts the skill. Plaintext is never
stored in the registry.

Each auditor claiming an audit task additionally stakes a
\emph{confidentiality bond}. Proven content leakage forfeits the bond
and resets the auditor's reputation to zero, which discourages
disclosure.

After completing the review, each auditor returns a
\textbf{\HeaderBlack{4} Signed Verdict.} The verdict is a signed report
containing a vote $v \in \{\texttt{safe}, \texttt{unsafe},
\texttt{abstain}\}$, a list of $\mathit{risk\_findings}$, and a
confidence score $c \in [0,1]$. The auditor submits the report to the
Voting Contract, which aggregates non-abstaining ballots into the
admission decision detailed in \S\ref{sec:design:dao}.

\subsection{Anchoring Stage}
\label{sec:design:dao}

The anchoring stage commits every approved skill, together with its
audit report and metadata, to the Skill Registry, and
settles auditor incentives by rewarding consensus-aligned voters
and slashing those whose votes diverged. It proceeds through two
steps: the Voting Contract aggregates non-abstaining ballots through
reputation-weighted voting and writes the resulting record to the
Skill Registry; the Reputation \& Staking Contract then applies the
corresponding reward or slash to each participating auditor.

Once the signed verdicts are received, the protocol performs a
\textbf{\HeaderBlueGray{5} Status Update.} Beyond updating the
audit-status flag, this step migrates every approved skill from the
Pre-Registry to the Skill Registry, where the skill becomes
invocable. The Voting Contract gates this promotion by aggregating
the non-abstaining verdicts through reputation-weighted voting, with
vote weight set in proportion to reputation alone, $w_i = r_i$,
where $r_i \in [0, 1000]$ is auditor $i$'s accumulated reputation
score, governed by the update rule in \S\ref{sec:design:reputation}.
We exclude stake from the weight formula because including stake would
let a capital-rich adversary dominate voting regardless of audit track
record; stake instead serves only as an anti-Sybil entry barrier and a
source of slash funds. The aggregate safety score is
\begin{equation}
  \mathit{SafeScore} =
    \frac{\sum\nolimits_{i \in \mathcal{A}_{\texttt{safe}}} w_i}
         {\sum\nolimits_{i \in \mathcal{A} \setminus \mathcal{A}_{\texttt{abstain}}} w_i},
  \label{eq:safe-score}
\end{equation}
where $\mathcal{A}$ is the set of participating auditors,
$\mathcal{A}_{\texttt{safe}}$ the subset voting \texttt{safe}, and
$\mathcal{A}_{\texttt{abstain}}$ the subset abstaining. The skill is
approved if and only if $\mathit{SafeScore} \geq \theta$, for a
governance-configurable threshold $\theta$ (\eg $\theta = 0.6$).
A minimum-auditor requirement of $N \geq 5$ further prevents a single
colluding auditor from unilaterally approving a malicious skill.

Once approved, the skill is migrated from the Pre-Registry to the
Skill Registry, a tamper-evident store that holds, for each accepted
skill, its content or cryptographic digest, permission manifest,
audit report, and versioning metadata. The audit report records the
participating auditors and the detailed audit result; entries in the
Skill Registry are by definition \texttt{approved}. A
$\mathit{prev\_version}$ pointer links each skill to its predecessor,
forming a chain of versions for full release traceability. Updates publish a new version: the developer submits
revised content, which produces a fresh $\mathit{skill\_id}$ and
re-enters the audit cycle, while the previously approved version
remains available until the new version passes review, so service
continues across transitions.

After the admission decision is finalized, the protocol applies a
\textbf{\HeaderBlueGray{6} Reward / Slash Update.}
\label{sec:design:reputation}
The audit outcome is forwarded to the Reputation \& Staking Contract,
which applies asymmetric rewards and penalties: stake settles
immediately at audit close, while reputation updates follow from the
skill's downstream behavior. We formalize the underlying economic
model and prove its incentive properties in
\S\ref{sec:design:economics}.

A new auditor enters the system by staking at least
$s_{\min}$ and receiving an initial reputation score $r_0 = 100$, a
conservative starting value that forces new auditors to build a track
record before exercising significant voting influence. Once a skill
survives a monitoring window $\Delta t_{\mathit{mon}}$ without incident,
every auditor who voted \texttt{safe} earns a positive reputation
increment $\delta^{+}$. Conversely, when a previously approved skill is later
proven malicious, the protocol penalizes auditors asymmetrically:
\begin{itemize}
\item Auditors who voted \texttt{safe} lose reputation
  ($r_i \leftarrow r_i - \delta^{-}$, with $\delta^{-} > \delta^{+}$)
  and forfeit a fixed slash amount $S_{\mathit{slash}}$ from their
  stake.
\item Auditors who correctly voted \texttt{unsafe} receive a
  reputation bonus and share the slashed tokens.
\end{itemize}

Two design choices keep this rule incentive-compatible:
$\delta^{-} > \delta^{+}$ prevents an attacker from offsetting one
proven bad vote with many trivial correct votes on benign skills,
and a fixed $S_{\mathit{slash}}$ (rather than stake-proportional)
deters auditors from staking only $s_{\min}$ to minimize exposure.

Reputation also undergoes slow time decay,
$r_i(t+1) = \alpha \cdot r_i(t)$ with $\alpha$ slightly below~1
(\eg $\alpha = 0.995$ per epoch). Decay prevents long-past audits
from shielding inactive auditors from the cost of newer mistakes,
and ties voting weight to continuous participation. Finally, when accumulated slashing reduces an auditor's
stake below $s_{\min}$, the auditor is deactivated until the stake is
replenished.

\subsection{Invocation Stage}
\label{sec:invocation}

The invocation stage releases every approved skill into the LLM
agent's context only through the SVL, a component integrated into
the AI Service Provider that serves both the LLM and the LLM
agent. The LLM or agent decides which $\mathit{skill\_id}$ or skill
name to invoke; the SVL decides whether and how to release the
corresponding artifact, querying the Skill Registry, decrypting the
content when necessary, and computing the effective permission
envelope before release. Because the SVL is the only loading path,
every skill that reaches the agent's context is, by construction,
audit-approved and intact, while any tampered skill is blocked.

The stage proceeds through six steps shared across all four
publication types, with type-specific branches at the access-check
(Step~\HeaderGreen{8}), payload-return (Step~\HeaderGreen{9}), and
local-integrity (Step~\HeaderGreen{10}) steps: the LLM forwards a
user's Skill Request to the SVL; the SVL submits a Skill Query to
the Skill Registry; the registry returns Hash, Ciphertext,
or Plaintext per publication type; the SVL performs Local Skill
Verification for Committed skills; the SVL runs Permission Check
over the relevant manifest(s); and the SVL releases the Verified
Skill into the LLM agent's context.

The invocation phase begins with a
\textbf{\HeaderGreen{7} Skill Request.} A user supplies a
$\mathit{skill\_id}$ or skill name, or a set of such identifiers when
the task needs multiple skills, optionally accompanied by a path to a
local file (for Committed skills), and the LLM agent forwards the
request to the SVL. Because the SVL is the sole entry point for skill
loading, the LLM may decide \emph{which} skill to load, but it cannot
decide \emph{whether} to verify or permission-check it, and a
previously loaded skill cannot trick the LLM into skipping either
control.

The request reaches the SVL, which issues a
\textbf{\HeaderGreen{8} Skill Query.}
\label{sec:design:verification}
The SVL queries the Skill Registry with the supplied $\mathit{skill\_id}$ or skill name.
All four publication types share two common preconditions: the skill
must be present in the Skill Registry (admitted as \texttt{approved}),
and the requesting user must satisfy the publication type's access
control. The branches differ only in
what the access check verifies:

\begin{itemize}
\item For a Transparent skill, no access control applies; any user with
  a valid $\mathit{skill\_id}$ or skill name may proceed.
\item For a Licensed skill, the SVL verifies that the delivery
  record for this user is non-empty, indicating that the developer has
  posted an encrypted content key for this buyer (wrapped via ECDH
  with the buyer's keypair, as detailed in Step~\HeaderGreen{9}).
\item For a Sealed skill, no access record is checked: the
  ciphertext is publicly downloadable, but only a holder of the
  developer-controlled private key $\mathit{sk}_d$ can decrypt it. Access is therefore
  proven implicitly through the decryption in Step~\HeaderGreen{9}.
\item For a Committed skill, possession of a local file matching the
  $\mathit{content\_hash}$ constitutes the access proof; the
  user supplies a path to this file, whose integrity is verified
  separately in Step~\HeaderGreen{10}.
\end{itemize}

If both preconditions hold, the registry returns the corresponding payload
as
\textbf{\HeaderGreen{9} Hash / Ciphertext / Plaintext.}
The payload format and the SVL's subsequent operation branch by publication
type:

\begin{itemize}
\item For a Transparent skill, the registry returns the plaintext directly.
  Content-addressing of the entry guarantees integrity, so the SVL
  performs no further cryptographic operation at this step.
\item For a Licensed skill, the registry returns the ciphertext.
  The SVL recovers the symmetric content key $k_{\mathit{content}}$ by
  computing the ECDH shared secret
  $S_b = \mathit{sk}_b \cdot \mathit{pk}_d$ from the buyer's
  private key $\mathit{sk}_b$ and the developer's public key
  $\mathit{pk}_d$, deriving a delivery key
  \[
    k_b = \mathrm{HKDF}(S_b,\; \texttt{license},\;
                        \mathit{skill\_id} \,\|\, \mathit{pk}_b \,\|\, \mathit{pk}_d),
  \]
  and unwrapping $k_{\mathit{content}}$ from the developer-posted
  record under $k_b$. It then decrypts the
   ciphertext under $k_{\mathit{content}}$ and verifies
  $H(\mathit{plaintext}) = \mathit{content\_hash}$. The \texttt{license}
  info tag domain-separates buyer delivery from auditor delivery in
  \S\ref{sec:design:publication} (info tag \texttt{audit}), so the same
  key pair derives independent delivery keys across the two contexts. We
  use ECDH with HKDF rather than a single symmetric key shared with every
  buyer because the developer never learns any buyer's private key, and
  per-buyer key derivation limits the impact of a buyer-side key
  compromise to that buyer alone.
\item {\sloppy For a Sealed skill, the registry returns the encrypted blob. The SVL
computes $S_{\mathit{seal}} = \mathit{sk}_d \cdot \mathit{pk}_d$
through an ECDH self-handshake\footnote{For
Sealed skills the developer is also the user (or an internal
organization), so we re-use the ECDH primitive with the developer's
own keypair as a deterministic single-party key derivation.} on the
developer's own keypair. It then derives a per-skill sealing key
using a domain-separated HKDF invocation,
\[
  k_{\mathit{seal}} =
  \mathrm{HKDF}(S_{\mathit{seal}},\; \texttt{sealed},\; \mathit{skill\_id}),
\]
where \texttt{sealed} is an explicit domain-separation tag and
$\mathit{skill\_id}$ is included in the HKDF info field. The SVL
decrypts the blob under $k_{\mathit{seal}}$ with AES-256-GCM and
verifies $H(\mathit{plaintext}) = \mathit{content\_hash}$. Successful
decryption confirms possession of $\mathit{sk}_d$ and thereby
constitutes the access proof for Sealed skills, with no separate
registry-side access check required. The self-handshake binds each per-skill key
to the developer's identity, and the domain-separated HKDF tag
prevents key reuse across audit, license, and sealed-skill
derivation contexts, without ever storing decryption keys in the registry.\par}
\item For a Committed skill, the registry returns only
  $\mathit{content\_hash}$; the plaintext is supplied locally by the
  user, and integrity verification is deferred to
  Step~\HeaderGreen{10}.
\end{itemize}

For a Committed skill, the SVL then performs
\textbf{\HeaderGreen{10} Local Skill Verification.} Given the user-supplied
path, the SVL computes $h' = H(\mathit{local\_content})$ and tests
$h' = \mathit{content\_hash}$. A mismatch indicates post-audit tampering of
the local file, and the SVL aborts the invocation. We restrict this step
to Committed skills because, for Transparent, Licensed, and Sealed skills,
the registry already holds an integrity-checked copy of the content
(plaintext for Transparent; hash-anchored ciphertext for Licensed and
Sealed), whereas a Committed skill's plaintext lives only on the
developer's device and must therefore be re-checked against the registry
hash at every load.

With both content and integrity established, the SVL performs
\textbf{\HeaderGreen{11} Permission Check} against the registry-held
Permission Manifest. For a single loaded skill, the check is:
\begin{equation}
  \mathit{declared\_tools} \subseteq \mathcal{T}_{\textit{user}}
  \;\wedge\;
  \mathit{data\_scope} \subseteq \mathcal{D}_{\textit{user}},
  \label{eq:permission-check}
\end{equation}
where $\mathcal{T}_{\textit{user}}$ and $\mathcal{D}_{\textit{user}}$ are
the tool set and data scope authorized for the current user. If the skill
requests permissions beyond the user's scope, the SVL prompts the user for
additional authorization, or aborts the invocation.

For a multi-skill invocation, let $K=\{s_1,\ldots,s_n\}$ be the set of
skills loaded for the same user request, and let each skill carry
$M_i=(\mathcal{T}_i,\mathcal{D}_i,\mathcal{B}_i)$ for declared tools,
data scopes, and behavior bounds. The SVL computes the common permission
envelope as the manifest intersection:
\begin{equation}{\small
  \mathcal{T}_{\cap}=\bigcap_{i=1}^{n}\mathcal{T}_i,\quad
  \mathcal{D}_{\cap}=\bigcap_{i=1}^{n}\mathcal{D}_i,\quad
  \mathcal{B}_{\cap}=\bigwedge_{i=1}^{n}\mathcal{B}_i .}
  \label{eq:permission-intersection}
\end{equation}
The intersection $(\mathcal{T}_{\cap},\mathcal{D}_{\cap},\mathcal{B}_{\cap})$
is injected into the LLM as the default executable permission envelope:
actions inside it may proceed without further escalation only if
$\mathcal{T}_{\cap}\subseteq\mathcal{T}_{\textit{user}}$ and
$\mathcal{D}_{\cap}\subseteq\mathcal{D}_{\textit{user}}$ also hold. If
the intersection itself exceeds the user's authorized scope, the excess
permissions require explicit user authorization, or the invocation is
aborted. The SVL also computes the non-common permissions
\begin{equation}{\small
  \mathcal{T}_{\Delta}=\left(\bigcup_{i=1}^{n}\mathcal{T}_i\right)\setminus\mathcal{T}_{\cap},
  \quad
  \mathcal{D}_{\Delta}=\left(\bigcup_{i=1}^{n}\mathcal{D}_i\right)\setminus\mathcal{D}_{\cap},}
  \label{eq:permission-delta}
\end{equation}
and marks them as requiring explicit user authorization before execution.
Thus, a permission exposed by only one loaded skill cannot silently become
available to the whole multi-skill context. This default-deny treatment of
$\mathcal{T}_{\Delta}$ and $\mathcal{D}_{\Delta}$ limits cross-skill
interaction attacks in which one skill attempts to induce the LLM to use
another skill's broader manifest.

Once the permission check succeeds, the SVL releases the
\textbf{\HeaderGreen{12} Verified Skill} by injecting the verified content
and the effective permission envelope into the LLM agent's context.
Permission validation lives in this common closing step rather than in the
per-type branches because the Permission Manifest is anchored to an immutable registry record for
every publication type, so the check is identical regardless of how the
content was retrieved or decrypted.

\subsection{Economic Model}
\label{sec:design:economics}

The economic model keeps \sysname's audit ecosystem sustainable and
secure over time. It does so in two ways: rewarding honest auditors so
that they keep participating, and steadily eroding the stake and
reputation of malicious or persistently inaccurate auditors until
they exit. The model is method-agnostic: an auditor running static
analysis, sandbox testing, or LLM-based semantic review earns the
same token credit (TC) for the same correct vote on the same skill, so method choice
is an internal economic decision of each auditor.

A single TC settles every flow in the protocol:
publication fees, audit rewards, auditor stakes, slashes,
confidentiality bonds, and Licensed-skill prices. The exchange rate
\begin{equation}
  1\;\mathrm{TC} \;=\; \kappa_{\mathit{tc}}\;\text{LLM tokens}
  \label{eq:tc-anchor}
\end{equation}
is set by the on-chain governance and ties the size-scaling term of the audit
reward (Eq.~\ref{eq:r-base}) to the system's main cost driver. Four
roles exchange TC: developer, auditor, user, and Protocol Treasury,
through the flows detailed below.

At publication, the developer prefunds an audit budget that the
contract distributes atomically at audit close. The reference reward
\begin{equation}{\small
  R_{\mathit{base}}(j) \;=\; \beta_0 + \beta_1\cdot \frac{L_j}{\kappa_{\mathit{tc}}},}
  \label{eq:r-base}
\end{equation}
combines a flat floor $\beta_0$ for method-independent overhead with
a size-scaling term $\beta_1\,L_j/\kappa_{\mathit{tc}}$ calibrated so
that token-heavy methods stay viable on long skills; $L_j$ is the
skill content length in LLM tokens, and both $\beta_0$ and $\beta_1$
are governance-set DAO parameters.
The publication fee
\begin{equation}
  C_{\mathit{pub}}(j) \;=\; N\cdot R_{\mathit{base}}(j),
  \label{eq:pub-fee}
\end{equation}
splits into a protocol-fee cut
$\varphi_{\mathit{proto}}\cdot N\cdot R_{\mathit{base}}(j)$ to
Treasury and a remainder
$(1-\varphi_{\mathit{proto}})\cdot N\cdot R_{\mathit{base}}(j)$
funding the audit pool. At audit close, the contract atomically pays
each consensus-aligned auditor
\begin{equation}
  R_i(j) \;=\;
  (1-\varphi_{\mathit{proto}})\cdot R_{\mathit{base}}(j)\cdot
  \frac{r_i}{r_{\max}}\cdot
  \mathbf{1}\bigl\{v_i = \mathit{cons}(j)\bigr\},
  \label{eq:reward}
\end{equation}
where $r_{\max} = 1000$ is the reputation cap from
\S\ref{sec:design:dao} and $\mathit{cons}(j)$ denotes the consensus
verdict on skill $j$. The contract also slashes each non-consensus
vote
\begin{equation}
  S_{\mathit{slash}}(j) \;=\; \gamma\cdot R_{\mathit{base}}(j),
  \qquad \gamma=2,
  \label{eq:slash}
\end{equation}
and refunds the residual pool to the developer.

Three calibration choices together enforce incentive compatibility:
$\gamma=2$ ensures $S_{\mathit{slash}}(j) > R_{\mathit{base}}(j)$, as
required by the analysis in \S\ref{sec:eval-rq4};
$\delta^{-}=2\delta^{+}$ (concretely $-30$ vs $+15$) makes reputation
loss exceed gain; and $r_0/r_{\max}=0.1$ caps a fresh identity's
per-vote reward at $10\%$ of $R_{\mathit{base}}(j)$ even when aligned
with consensus, while a single non-consensus vote forfeits the full
$\gamma\cdot R_{\mathit{base}}(j)$. A one-shot game analysis in
\S\ref{sec:eval-rq4-game} shows that, under this calibration,
the resulting game has a unique Nash equilibrium, even when
developers can offer bribes.

\emph{Sybil resistance} follows from the asymmetric payoff. A bloc of
fresh identities earns at most $0.1\,R_{\mathit{base}}(j)$ per
aligned vote but forfeits $\gamma \cdot R_{\mathit{base}}(j)$ per
non-consensus vote, regardless of stake size, so stacking the DAO with
Sybil identities is unprofitable.

\emph{Retrospective slashing} handles skills later shown malicious
via third-party evidence or a stakeholder arbitration submission that
triggers a re-audit. If the re-audit reverses the original verdict,
  the approving auditors are slashed, and the retrospective slash pool $\Sigma=\sum_{i \in \mathcal{A}_{\texttt{safe}}} S_{\mathit{slash}}(j)$ splits as $0.3\,\Sigma$ to the whistleblower, $0.4\,\Sigma$ to dissenting auditors, and $0.3\,\Sigma$ to the Protocol Treasury.
The pool splits three ways: dissenting auditors who correctly voted
\texttt{unsafe} share $40\%$ as a reward for the correct minority
vote; the whistleblower receives $30\%$ to compensate the cost of
gathering evidence; and the remaining $30\%$ replenishes the Protocol
Treasury. Honest dissenters may not fully recover their immediate
slash from this TC share alone when the originally approving group
is small; the subsequent reputation bonus complements the monetary
compensation. Capping the dissenters' share below $100\%$ removes the
incentive to delay evidence until that share is large enough. If a
re-audit confirms the original verdict, the submitter pays the
re-audit fee with no payout, which deters frivolous challenges.

Licensed-skill monetization runs an escrow flow in parallel: the
developer freezes a delivery bond $B_{\mathit{deliver}}$ at
publication and sets a per-access price $P$. A user pays
$\mathit{Total} = P\,(1+\varphi_{\mathit{proto}})$, split $P$ to
developer and $\varphi_{\mathit{proto}}P$ to Treasury. If the
developer fails to post the encrypted key within
$\tau_{\mathit{deliver}}=24\,\mathrm{hours}$ of purchase,
$B_{\mathit{deliver}}$ is forfeit, the user receives a full refund
of $P$, and the remainder $B_{\mathit{deliver}}-P$ flows to Treasury.

The Protocol Treasury is seeded with an initial allocation $T_0$ minted
at protocol launch and is refilled by the protocol-fee cut, immediate
slashes from auditors whose votes diverged from consensus, the $30\%$
retrospective-slash share, and forfeited delivery bonds. Treasury outflows fund dispute
arbitration, contract upgrades, an emergency reserve, and a slippage
subsidy that raises the reference reward when fewer than $N$
auditors claim a pending skill within the base wait interval
$\Delta t$:
\begin{equation}
  R_{\mathit{base}}(j, t) \;=\;
    R_{\mathit{base}}(j)\cdot
    \Bigl(1 + \sigma\,\bigl\lfloor\tfrac{t-t_0}{\Delta t}\bigr\rfloor\Bigr),
  \label{eq:slippage}
\end{equation}
with a slippage rate $\sigma = 0.25$ per interval, ensuring that
unpopular-but-legitimate skills (\eg those requiring specialized
auditor expertise) still attract reviewers. All system parameters
($T_0$, $\varphi_{\mathit{proto}}$, $\beta_0$, $\beta_1$, $\gamma$,
$\sigma$, slash split) are subject to DAO vote.

\section{Evaluation}
\label{sec:eval}

We organize the evaluation around four research questions (RQs):
\begin{itemize}
  \item \textbf{RQ1.} Does \sysname cover the six representative attacks
    of \S\ref{sec:threat-model}, and how does it compare with existing
    defenses?
  \item \textbf{RQ2.} What runtime and economic overhead does \sysname
    incur?
  \item \textbf{RQ3.} How effective is each pluggable audit method
    standalone, and what does the DAO committee gain by aggregating them?
  \item \textbf{RQ4.} Does the economic mechanism of
    \S\ref{sec:design:economics} make honest auditing
    incentive-compatible?
\end{itemize}

\subsection{Setup and Implementation}
\label{sec:eval-setup}
\label{sec:impl}

We implement \sysname's governance and economic components in Solidity, and
the SVL and supporting components in Python. Four core
components realize \sysname:

\begin{itemize}
    \item \textbf{Token Credit Contract}: the contract that issues the TC of \S\ref{sec:design:economics}, used for auditor stakes, rewards, and slashes.
    \item \textbf{Skill Pre-Registry} and \textbf{Skill Registry}: the former records pending submissions and audit requests; the latter stores, for each approved skill, its content or cryptographic digest, permission manifest, audit report, and versioning metadata across all four publication types. We implemented the registry over four backends: Ethereum Sepolia, immudb, Tessera, and Solana.
    \item \textbf{Voting Contract}: aggregates signed auditor verdicts through reputation-weighted DAO voting and gates promotion from the Skill Pre-Registry to the Skill Registry.
    \item \textbf{Reputation \& Staking Contract}: manages auditor reputation, stake deposits, and reward/slash updates.
\end{itemize}

\sysname provides six pluggable audit methods that auditors run
independently: \emph{Static}~\cite{jamshidi2025securing}, \emph{LLM},
\emph{SkillScan}~\cite{liu2026agentskillswild},
\emph{Manifest}~\cite{shi2025progent},
\emph{Repello SkillCheck}~\cite{repello2026skillcheck}, and
\emph{Security Scanner}~\cite{mcpmarket2026scanner}. The SVL is implemented as a Python module on the
agent's loading path.

The dataset used in this evaluation contains 1{,}079 real-world skills collected from ClawHub,
GitHub repositories, and other public skill registries. We label each
skill through automated security scanning~\cite{snyk2026toxicskills},
metadata analysis, and manual review. Automated findings serve only as
candidate evidence: the final label follows a written attack-category
rubric and manual inspection of the skill's instructions, metadata,
scripts, and declared permissions. Ambiguous cases are discussed until
the reviewers reach consensus rather than being assigned by scanner
majority. This procedure yields 779 malicious and 300 benign skills; 93
of the malicious skills correspond to
\textbf{explicit injection} or \textbf{implicit poisoning}.
Because scanner outputs informed candidate selection and some evaluated
auditors are scanner-based, the labels are not fully independent of all
evaluated methods. We therefore treat possible incorporation and
distribution bias as a threat to validity rather than interpreting the
reported accuracy as performance on an independently sampled population.
To our knowledge, this corpus is among
the most comprehensive skill-security benchmarks to date.

Appendix~\ref{sec:param-experiments} reports
sensitivity sweeps for the initial reputation ratio
$r_0/r_{\max}$, the approval threshold $\theta$, the number of
audit methods, and the slash coefficient $\gamma$, and justifies
the values used throughout this section.

\subsection{Security Analysis}
\label{sec:eval-analysis}

\begin{table}[t]
\definecolor{rqBlue}{RGB}{42,94,171}
\definecolor{rqRed}{RGB}{180,45,45}
\newcommand{\rqFullMark}{\begingroup\large\ding{51}\endgroup}
\newcommand{\rqMechMark}{\begingroup\large$\blacklozenge$\endgroup}
\newcommand{\rqHalfMark}{\begingroup\large$\bigcirc$\endgroup}
\newcommand{\rqEmptyMark}{\begingroup\large\ding{55}\endgroup}
\newcommand{\rqPad}{\vphantom{\textcolor{rqBlue}{\small\bfseries 100\%}}}
\newcommand{\rqPctOffset}{0.78em}
\newcommand{\rqCell}[1]{\makebox[\linewidth][l]{\hspace*{0.24\linewidth}#1}}
\newcommand{\rqFull}{\rqCell{\textcolor{rqBlue}{\rqFullMark}\rqPad}}
\newcommand{\rqHalf}{\rqCell{\rqHalfMark\rqPad}}
\newcommand{\rqEmpty}{\rqCell{\rqEmptyMark\rqPad}}
\newcommand{\rqNA}{\rqCell{\textemdash\rqPad}}
\newcommand{\rqPct}[3]{\rqCell{\makebox[0pt][c]{#1}\makebox[0pt][l]{\hspace{\rqPctOffset}\textcolor{#2}{\small\bfseries #3}}\rqPad}}
\newcommand{\rqFullPct}[1]{\rqPct{\textcolor{rqBlue}{\rqFullMark}}{rqBlue}{#1}}
\newcommand{\rqHalfPct}[1]{\rqPct{\rqHalfMark}{rqBlue}{#1}}
\newcommand{\rqMechPct}[1]{\rqPct{\textcolor{rqRed}{\rqMechMark}}{rqRed}{#1}}
\newcommand{\rothead}[1]{\rotatebox[origin=lB]{90}{\textbf{#1}}}
\newcommand{\rqMechNote}{\textsuperscript{\textcolor{rqBlue}{\dag}}}
\newcommand{\rotheadtwo}[2]{\rotatebox[origin=lB]{90}{\shortstack[l]{\textbf{#1}\\\textbf{#2}}}}
\newcolumntype{C}{>{\centering\arraybackslash}m{4.0em}}
\caption{Resistance of \sysname and six baselines against the six
representative attacks of \S\ref{sec:threat-model}.}
\label{tab:rq1-defense-comparison}
\centering
\normalsize
\setlength{\tabcolsep}{4pt}
\setlength{\arrayrulewidth}{0.4pt}
\renewcommand{\arraystretch}{1.28}
\rowcolors{0}{gray!12}{white}
\resizebox{\columnwidth}{!}{%
\begin{tabular}{l|CCCC|CC}
\toprule
\rowcolor{white}
\textbf{Defense} & \multicolumn{4}{c|}{\textbf{Pre-anchoring attacks}} & \multicolumn{2}{c}{\textbf{Post-anchoring}} \\
\cmidrule(lr){2-5}\cmidrule(lr){6-7}
\rowcolor{white}
 & \rotheadtwo{Explicit}{injection}
 & \rotheadtwo{Implicit}{poisoning}
 & \rotheadtwo{Cross-skill}{interaction}
 & \rotheadtwo{Auditor}{collusion}
 & \rothead{Rug pull}
 & \rotheadtwo{Local}{tampering} \\
\midrule
SkillScan~\cite{liu2026agentskillswild}   & \rqFullPct{86.7\%}  & \rqFullPct{70.5\%}  & \rqEmpty & \rqNA   & \rqHalf  & \rqHalf \\
SkillProbe~\cite{guo2026skillprobe}           & \rqFullPct{90\%}  & \rqFullPct{86\%}  & \rqFullPct{87\%}  & \rqNA   & \rqHalf  & \rqHalf \\
MalSkills~\cite{wang2026malskills}  & \rqFullPct{95\%}  & \rqFullPct{85\%}  & \rqEmpty & \rqNA   & \rqHalf  & \rqHalf \\
AgentFuzz\cite{liu2025make}            & \rqEmpty & \rqEmpty & \rqFullPct{97.14\%}  & \rqNA   & \rqFullPct{92.8\%}  & \rqEmpty \\
Mainstream markets~\cite{hou2025security}       & \rqHalf  & \rqHalf  & \rqEmpty & \rqNA   & \rqEmpty  & \rqEmpty \\
AgentSentinel~\cite{hu2025agentsentinel}                & \rqEmpty & \rqEmpty & \rqFullPct{87\%}  & \rqNA   & \rqHalf  & \rqEmpty \\
\textbf{\sysname}                            & \rqFullPct{97.6\%}  & \rqFullPct{97.6\%}  & \rqFullPct{90.2\%}  & \rqFullPct{40\%\rqMechNote} & \rqFullPct{95.1\%}  & \rqMechPct{100\%} \\
\bottomrule
\end{tabular}%
}
\vspace{2pt}
\parbox{\linewidth}{\footnotesize
\textcolor{rqBlue}{\rqFullMark}{} empirical-level resists; \textcolor{rqRed}{\rqMechMark}{} mechanism-level guarantee; $\bigcirc$ partially resists;
\rqEmptyMark{} does not resist; \textemdash{} not applicable.
\textcolor{rqBlue}{40\%\rqMechNote} is the highest tolerable ratio of
malicious auditors.}
\end{table}

We argue that \sysname's design mitigates the six representative
attacks of Section~\ref{sec:threat-model}, assuming an honest majority
of auditor voting power.


For \textbf{explicit injection} and \textbf{implicit poisoning}, the
attack succeeds only if a harmful skill passes DAO
audit. \sysname requires independent auditors, aggregates their verdicts
by reputation-weighted consensus (Equation~\ref{eq:safe-score}), and
records signed verdicts for later accountability. The stake-and-slash
mechanism (\S\ref{sec:design:economics}) makes dishonest approval costly
and enables retrospective punishment when an approved skill is later shown
malicious.

For \textbf{rug pull}, the modification is introduced on the registry-side.
Through the registry backend's tamper-evident constraints, an approved skill can only be
updated through a new DAO audit process. For \textbf{local tampering},
\sysname relies on content-addressed skill identifiers and SVL hash checks.
Any modification after approval changes the recomputed content hash, so
the SVL rejects the artifact before it enters the agent context
(\S\ref{sec:design:verification}). This covers all four publication
types: Transparent plaintext is anchored to the immutable registry record;
Licensed and Sealed plaintext is verified after decryption against the
content hash in that record (Step~\HeaderGreen{9}); and Committed plaintext
is re-hashed against the immutable registry record at every load
(Step~\HeaderGreen{10}).

For \textbf{cross-skill interaction}, \sysname bounds the permissions
available to a multi-skill context rather than treating loaded skills as
a permission union. During invocation, the SVL performs
Permission Check (Step~\HeaderGreen{11}) over the loaded skills' manifests
(\S\ref{sec:design:verification}): only the manifest intersection that
also lies within the user's authorized scope is injected as the default
executable envelope, while permissions in the union-minus-intersection
set require explicit user authorization or cause the invocation to
abort. A malicious skill therefore cannot silently induce the LLM to use
another loaded skill's broader tool or data permissions. For
\textbf{auditor collusion}, stake, reputation-weighted voting, slashing,
and signed audit records make coordinated dishonest approval accountable.

\begin{figure}[t]
  \centering
  \includegraphics[width=0.8\columnwidth]{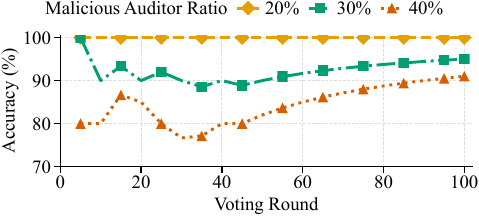}
  \caption{Audit accuracy over voting rounds under dynamic
  reputation updates.}
  \Description{Skill audit accuracy is plotted over voting rounds under dynamic reputation updates for malicious auditor thresholds of 20 percent, 30 percent, and 40 percent.}
  \label{fig:collusion}
\end{figure}

\subsection{Defense Effectiveness}
\label{sec:eval-rq1}

{\sloppy
End-to-end resistance is the strongest property a skill-security
framework can claim: regardless of how an attacker enters the supply
chain, the framework should keep harmful content out of the agent's
context. We therefore evaluate \sysname against the six representative
attacks of \S\ref{sec:threat-model} and compare it with six external
defenses: SkillScan~\cite{liu2026agentskillswild},
SkillProbe~\cite{guo2026skillprobe},
MalSkills~\cite{wang2026malskills},
AgentFuzz~\cite{liu2025make},
mainstream marketplace review~\cite{hou2025security}, and
AgentSentinel~\cite{hu2025agentsentinel}.
\par}
Explicit injection and implicit poisoning use the 93 malicious
skill-content samples from the corpus; rug pull and local tampering
use 250 tampered artifacts derived from 50 clean skills; cross-skill
interaction uses 33 malicious skills; and auditor collusion is
evaluated through a discrete-event voting simulation. We classify
each defense as fully resistant, partially resistant, not resistant,
or not applicable to a given attack.
Table~\ref{tab:rq1-defense-comparison} reports the results;
full-corpus audit quality is evaluated separately in
\S\ref{sec:eval-rq3}.

For \textbf{explicit injection} and \textbf{implicit poisoning}, we use
the same 93 malicious skill-content samples for both attacks because
they jointly cover overt instructions to access credentials, invoke
undeclared tools, or exfiltrate
results~\cite{greshake2023indirect,schulhoff2023hackaprompt}, as well
as semantically disguised descriptions and metadata that bias tool
selection~\cite{mo2026attractive,wang2025mcptox}. \sysname reaches 97.6\%
accuracy on this set by combining semantic review with
permission-manifest checks. Content-auditing baselines (SkillScan,
SkillProbe, MalSkills) also perform well when attacks leave lexical,
structural, or semantic evidence at submission, whereas runtime-only
defenses and marketplace review do not target submission-time content.

\begin{figure*}[t]
  \centering
  \begin{minipage}[t]{0.22\linewidth}
    \centering
    \includegraphics[width=\linewidth]{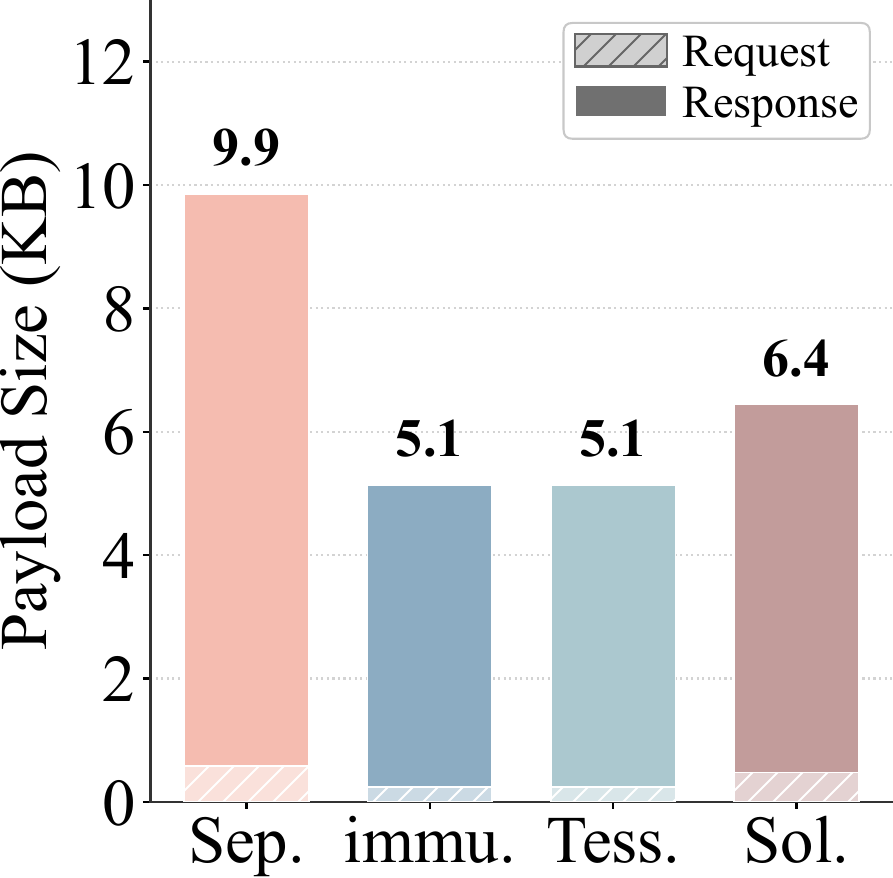}
    \vfill
    \caption{Skill payload size across the four backends.}
    \Description{Request, response, and total payload sizes are compared across Sepolia, immudb, Tessera, and Solana.}
    \label{fig:network-overhead}
  \end{minipage}
  \hfill
  \begin{minipage}[t]{0.22\linewidth}
    \centering
    \includegraphics[width=\linewidth]{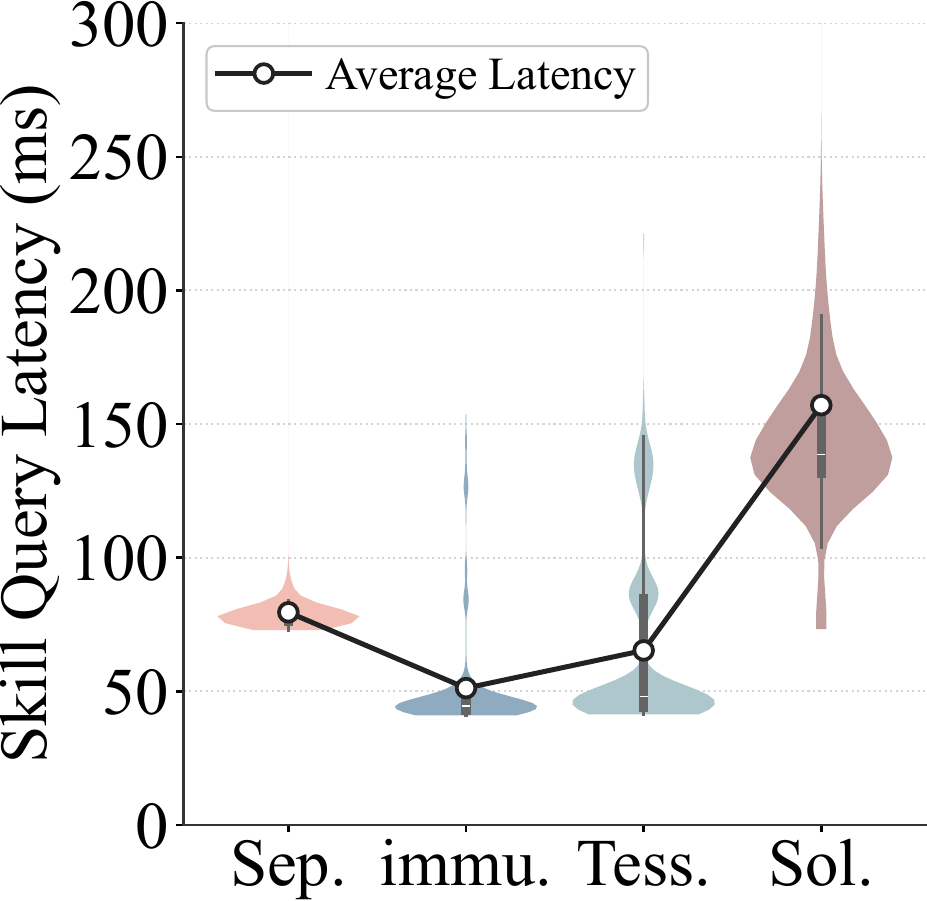}
    \vfill
    \caption{Per-round skill query latency across the four backends.}
    \Description{The distributions and average skill query latency are compared across Sepolia, immudb, Tessera, and Solana.}
    \label{fig:latency-rounds}
  \end{minipage}
  \hfill
  \begin{minipage}[t]{0.22\linewidth}
    \centering
    \includegraphics[width=\linewidth]{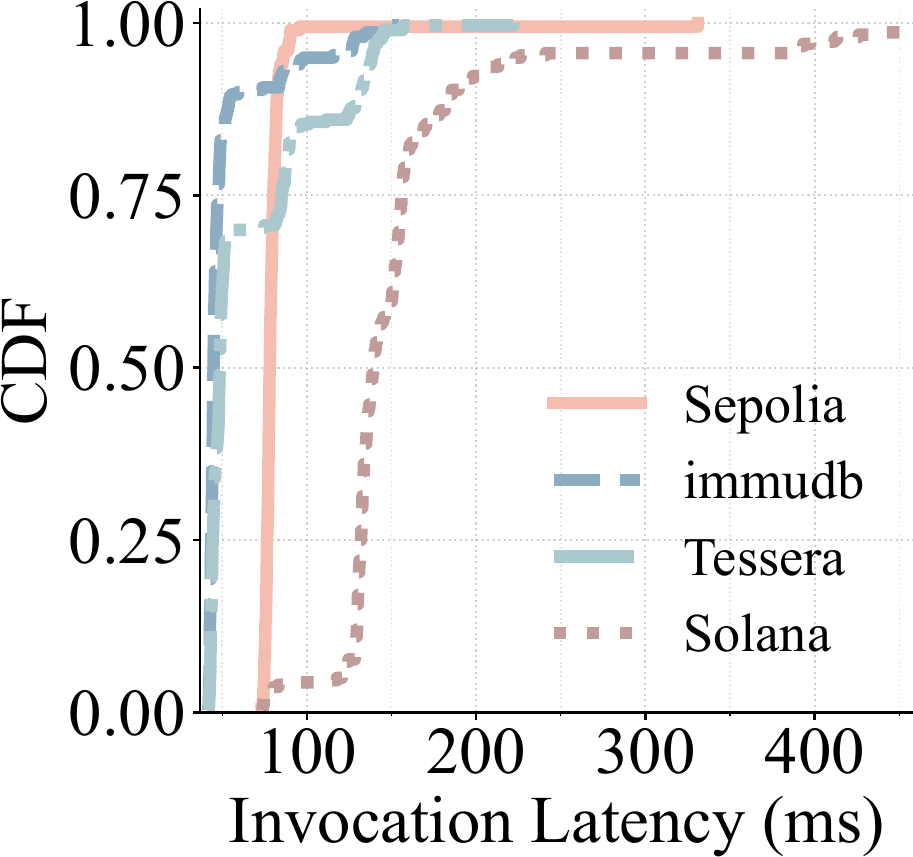}
    \vfill
    \caption{CDFs of invocation latency across the four backends.}
    \Description{Empirical cumulative distributions of invocation latency are compared across Sepolia, immudb, Tessera, and Solana.}
    \label{fig:latency-cdf}
  \end{minipage}
  \hfill
  \begin{minipage}[t]{0.22\linewidth}
    \centering
    \includegraphics[width=\linewidth]{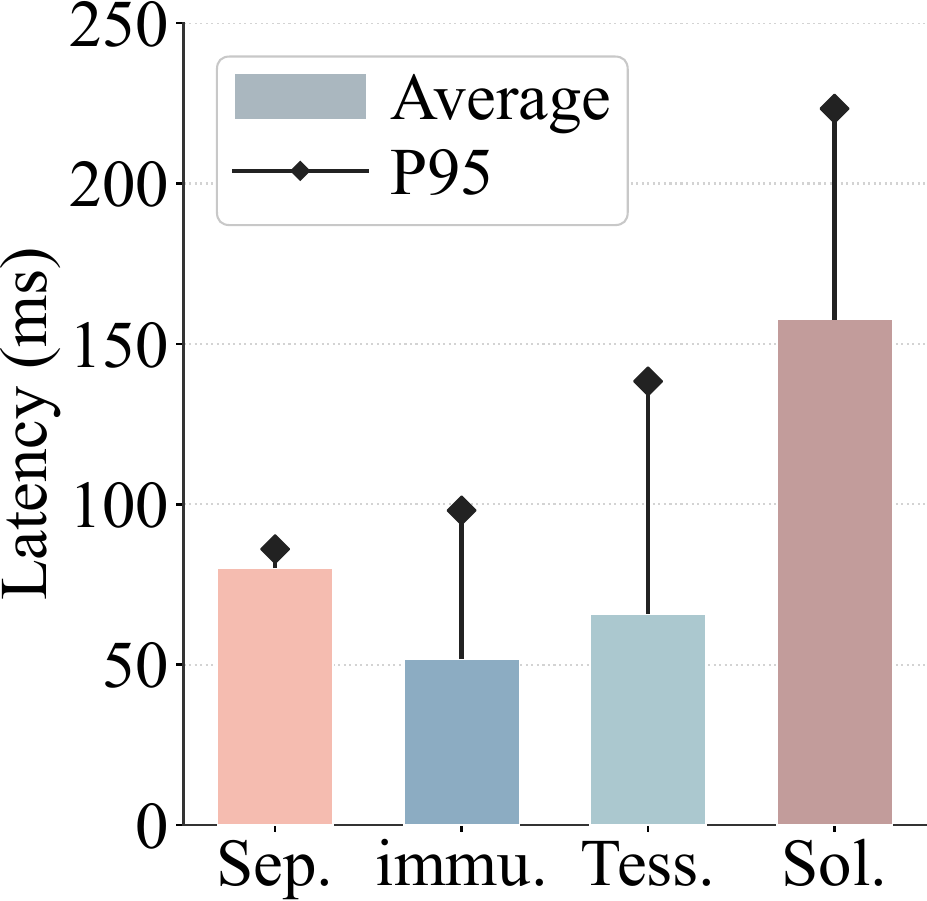}
    \vfill
    \caption{Average and P95 invocation latency across the four backends.}
    \Description{Average and P95 invocation latency are compared across Sepolia, immudb, Tessera, and Solana.}
    \label{fig:latency-p95}
  \end{minipage}
\end{figure*}

\textbf{Rug pull} is evaluated by approving 50 clean skills and
applying five post-approval tampering patterns, yielding 250
attempts~\cite{jamshidi2025securing,song2025beyond}. Because \sysname stores approved
skills as tamper-evident registry versions, an approved artifact cannot be silently replaced.
When the 250 modified variants are instead submitted as new versions, the DAO audit pipeline rejects 95.1\% of them before admission. Although
re-auditing and runtime testing may detect some modifications, they do
not cryptographically bind the audited artifact to the artifact later
loaded by the agent.

\textbf{Cross-skill interaction} is evaluated on 33 malicious skills
covering shadowing, Puppet attacks, permission aggregation, and
output injection~\cite{song2025beyond,zong2025mcpsafetybench}.
\sysname mitigates 90.2\% of these attacks through the Permission Check
argued in \S\ref{sec:eval-analysis}. The one missed case stays
within the user-authorized manifest intersection, which marks the
limit of manifest-based checks. This evaluation supports only
a permission-level mitigation claim; it does not establish defense against
semantic collusion through shared context, outputs, naming conventions, or
data formats. Baselines that reason across skills
or intercept runtime actions provide partial to strong coverage;
single-skill static review does not capture this class.

\textbf{Auditor collusion} targets governance rather than skill content.
To evaluate, we simulate 20 auditors
over 100 voting rounds with coordinated malicious auditors at 20\%, 30\%,
and 40\%. Figure~\ref{fig:collusion} shows that dynamic reputation and
slashing improve audit accuracy across all three malicious-auditor ratios.

\textbf{Local tampering} reuses the 250 rug-pull artifacts as
user-side modifications to cached or committed
skills~\cite{li2026towards}. \sysname blocks every attempt (100\%
accuracy) by comparing the local hash with the registry record on
the default loading path. Static and semantic scanners help only if
the user re-scans manually; runtime defenses and marketplace review
do not verify local integrity before loading.

\begin{summarybox}
\noindent\textbf{Answer to RQ1:}\quad \sysname is the only system that provides coverage across all six
representative attacks and maintains stable accuracy under
20\%--40\% malicious-auditor collusion.
\end{summarybox}

\begin{table}[b]
\caption{Audit token usage on a 100-skill subsample. Quota Share
is the share of a $15{,}000$K-token GPT subscription quota.}
\label{tab:audit-token-cost}
\centering
\small
\setlength{\tabcolsep}{3.0pt}
\renewcommand{\arraystretch}{1.1}
\newcolumntype{A}{>{\columncolor{gray!8}[0pt][\tabcolsep]}l}
\newcolumntype{B}{>{\columncolor{gray!8}[\tabcolsep][\tabcolsep]\centering\arraybackslash}m{0.42\linewidth}}
\newcolumntype{C}{>{\columncolor{gray!8}[\tabcolsep][\tabcolsep]\centering\arraybackslash}c}
\newcolumntype{D}{>{\columncolor{gray!8}[\tabcolsep][0pt]\centering\arraybackslash}c}
\resizebox{\linewidth}{!}{%
\begin{tabular}{@{}l>{\centering\arraybackslash}m{0.42\linewidth}cc@{}}
\toprule
\textbf{Method} &
\textbf{Cost distribution} &
\textbf{Avg. Tokens} &
\textbf{Quota Share} \\
& & (K tokens) & (\%) \\
\midrule
\multicolumn{1}{@{}>{\columncolor{gray!8}[0pt][\tabcolsep]}l}{Static~\cite{jamshidi2025securing}}
  & \multicolumn{1}{>{\columncolor{gray!8}[\tabcolsep][\tabcolsep]\centering\arraybackslash}m{0.42\linewidth}}{\textemdash}
  & \multicolumn{1}{>{\columncolor{gray!8}[\tabcolsep][\tabcolsep]\centering\arraybackslash}c}{0.00}
  & \multicolumn{1}{>{\columncolor{gray!8}[\tabcolsep][0pt]\centering\arraybackslash}c@{}}{0.000} \\
LLM          & \includegraphics[width=\linewidth]{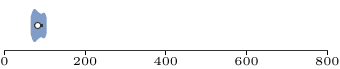} & 84.02  & 0.560 \\
\multicolumn{1}{@{}>{\columncolor{gray!8}[0pt][\tabcolsep]}l}{SkillScan~\cite{liu2026agentskillswild}}
  & \multicolumn{1}{>{\columncolor{gray!8}[\tabcolsep][\tabcolsep]\centering\arraybackslash}m{0.42\linewidth}}{\includegraphics[width=\linewidth]{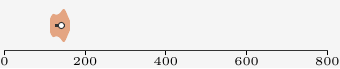}}
  & \multicolumn{1}{>{\columncolor{gray!8}[\tabcolsep][\tabcolsep]\centering\arraybackslash}c}{137.70}
  & \multicolumn{1}{>{\columncolor{gray!8}[\tabcolsep][0pt]\centering\arraybackslash}c@{}}{0.918} \\
Manifest~\cite{shi2025progent}     & \includegraphics[width=\linewidth]{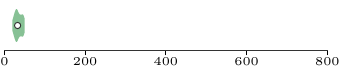} & 34.87  & 0.232 \\
\multicolumn{1}{@{}>{\columncolor{gray!8}[0pt][\tabcolsep]}l}{Security Scanner~\cite{mcpmarket2026scanner}}
  & \multicolumn{1}{>{\columncolor{gray!8}[\tabcolsep][\tabcolsep]\centering\arraybackslash}m{0.42\linewidth}}{\includegraphics[width=\linewidth]{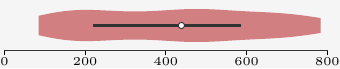}}
  & \multicolumn{1}{>{\columncolor{gray!8}[\tabcolsep][\tabcolsep]\centering\arraybackslash}c}{418.79}
  & \multicolumn{1}{>{\columncolor{gray!8}[\tabcolsep][0pt]\centering\arraybackslash}c@{}}{2.792} \\
\bottomrule
\end{tabular}
}
\end{table}

\subsection{System Overhead}
\label{sec:eval-rq2}

A defense is useful only if its overhead is acceptable. Two
overheads matter for \sysname: the LLM tokens consumed by auditing
(an economic cost paid once per submission) and the latency added by
the SVL on the loading path (a runtime cost paid on every
invocation). We measure both.

Audit-token usage is sampled on a 100-skill subset and normalized
against a $15{,}000$K-token monthly GPT quota for a \$20 subscription
plan, taken as a conservative reference for typical paid users.
Static is rule-based and uses no LLM tokens;
Repello~\cite{repello2026skillcheck} is excluded because its hosted
interface does not expose per-request token usage. For loading, the SVL issues one registry request
per skill and evaluates both backend query latency and end-to-end invocation latency across Ethereum Sepolia, immudb, Tessera, and Solana.
Each backend contributes 100 measured runs.

Table~\ref{tab:audit-token-cost} reports the token-usage results:
Manifest has the lowest LLM cost (34.87K tokens, 0.232\%), followed
by LLM (84.02K, 0.560\%) and SkillScan (137.70K, 0.918\%). Security
Scanner is highest (418.79K, 2.792\%) because its script-derived
findings require LLM normalization. Every measured method stays
below 3\% of the \$20-plan quota.


Figures~\ref{fig:network-overhead}--\ref{fig:latency-p95} report the skill-query and invocation results. Total retrieval payload
is 9.9 KB on Sepolia, 5.1 KB on immudb and Tessera, and 6.4 KB
on Solana. Mean (P95) query latency is 79.6 (85.4) ms on Sepolia,
51.2 (97.6) ms on immudb, 65.3 (137.9) ms on Tessera, and 157.1
(222.9) ms on Solana. Including SVL validation yields mean (P95) invocation
latencies of 80.1 (86.0), 51.7 (98.1), 65.8 (138.3), and 157.6
(223.4) ms, respectively, as shown in Figure~\ref{fig:latency-p95}. Invocation latency
covers backend retrieval and SVL validation; the latter is performed locally by the SVL according to the publication type.

To further characterize the SVL validation component, we select a
13{,}395-byte benign skill at the median of the benign-skill size
distribution in our dataset. Over 100 runs, mean validation latency
is 0.041\,ms for Transparent, 0.172\,ms for Committed,
0.495\,ms for Licensed, and 0.438\,ms for Sealed. Transparent
performs permission and integrity checks; Committed additionally
reads the local skill; Licensed and Sealed further recover or derive
the decryption key, decrypt the skill, and verify its integrity.

\begin{summarybox}
\noindent\textbf{Answer to RQ2:}\quad The measured overhead is minor: audit token usage stays under 3\% of a \$20-plan monthly quota, while mean skill-query latency ranges from 51.2 to 157.1\,ms across the four backends. Local SVL validation adds at most 0.495 ms across the four publication types.
\end{summarybox}

\subsection{Audit Effectiveness}
\label{sec:eval-rq3}

\begin{figure*}[t]
  \centering
  \includegraphics[width=\textwidth]{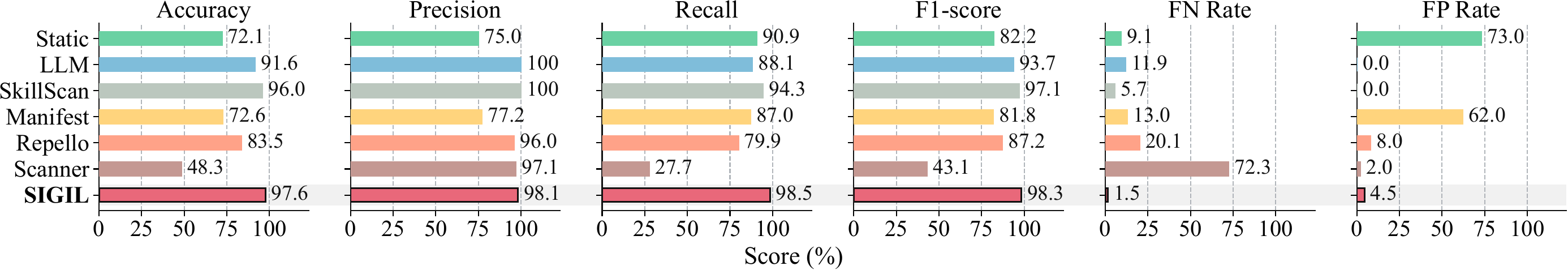}
  \caption{Aggregate audit metrics across six skill audit methods
  and DAO Committee voting.}
  \Description{A six-panel horizontal bar chart compares accuracy,
  precision, recall, F1-score, FN rate, and FP rate for the six skill audit
  methods and DAO Committee voting, showing SkillScan and the committee as the
  strongest overall configurations.}
  \label{fig:rq3-metrics-six-panel}
\end{figure*}

\begin{table}[b]
\caption{Leave-one-out ablation of the six audit methods. Each row
removes one audit method from the DAO committee; metrics in percent.}
\label{tab:rq3-auditor-ablation}
\centering
\small
\setlength{\tabcolsep}{3.5pt}
\renewcommand{\arraystretch}{1.12}

\newcommand{\drop}[2]{#1{\scriptsize\textcolor{red}{\,$\blacktriangledown$\,#2}}}
\newcommand{\rise}[2]{#1{\scriptsize\textcolor{blue}{\,$\blacktriangle$\,#2}}}
\newcommand{\gain}[2]{#1{\scriptsize\textcolor{blue}{\,$\blacktriangle$\,#2}}}
\newcommand{\betterlow}[2]{#1{\scriptsize\textcolor{red}{\,$\blacktriangledown$\,#2}}}
\rowcolors{0}{gray!12}{white}

\resizebox{\columnwidth}{!}{%
\begin{tabular}{@{}lrrrrrr@{}}
\toprule
\rowcolor{white}
\textbf{Variant} &
\textbf{Accuracy} &
\textbf{Precision} &
\textbf{Recall} &
\textbf{F1-score} &
\textbf{FN Rate} &
\textbf{FP Rate} \\
\midrule
W/o Static
& \drop{95.64}{1.98}
& \gain{99.96}{1.83}
& \drop{93.88}{4.64}
& \drop{96.82}{1.50}
& \rise{6.12}{4.64}
& \betterlow{0.10}{4.42} \\

W/o LLM
& \drop{95.00}{2.62}
& \drop{98.06}{0.07}
& \drop{94.80}{3.72}
& \drop{96.40}{1.92}
& \rise{5.20}{3.72}
& 4.52{\scriptsize\textcolor{gray}{\,--}} \\

W/o SkillScan
& \drop{93.82}{3.80}
& \drop{98.02}{0.11}
& \drop{93.14}{5.38}
& \drop{95.52}{2.80}
& \rise{6.86}{5.38}
& 4.52{\scriptsize\textcolor{gray}{\,--}} \\

W/o Manifest
& \drop{96.23}{1.39}
& \gain{99.95}{1.82}
& \drop{94.71}{3.81}
& \drop{97.26}{1.06}
& \rise{5.29}{3.81}
& \betterlow{0.12}{4.40} \\

W/o Repello SkillCheck
& \drop{96.56}{1.06}
& \gain{99.61}{1.48}
& \drop{95.51}{3.01}
& \drop{97.51}{0.81}
& \rise{4.49}{3.01}
& \betterlow{0.91}{3.61} \\

W/o Security Scanner
& \drop{97.42}{0.20}
& \gain{98.49}{0.36}
& \drop{97.85}{0.67}
& \drop{98.17}{0.15}
& \rise{2.15}{0.67}
& \betterlow{3.62}{0.90} \\

\midrule
\rowcolor{white}
\textbf{\sysname}
& \textbf{97.62}
& \textbf{98.13}
& \textbf{98.52}
& \textbf{98.32}
& \textbf{1.48}
& \textbf{4.52} \\
\bottomrule
\end{tabular}%
}
\vspace{2pt}
\parbox{\linewidth}{\footnotesize
\textcolor{red}{$\blacktriangledown$}: decrease;
\textcolor{blue}{$\blacktriangle$}: increase, relative to full
\sysname.}
\end{table}

\sysname's audit framework is pluggable: any auditor that joins the
DAO can contribute. We evaluate both
halves of this design --- how each method performs standalone and how
much the DAO Committee gains by aggregating them. This experiment
evaluates the effectiveness of method aggregation, not the accuracy of
the deployed DAO protocol.

We benchmark six audit methods on the 1{,}079-skill corpus
(779 malicious, 300 benign): \textbf{Static}, \textbf{LLM},
\textbf{SkillScan}~\cite{liu2026agentskillswild}, \textbf{Manifest},
\textbf{Repello SkillCheck}~\cite{repello2026skillcheck}, and
\textbf{Security Scanner}~\cite{mcpmarket2026scanner}. For each method
we compute a binary confusion matrix and report accuracy, precision,
recall, F1-score, FN rate, and false-positive (FP) rate. The evaluated
committee is an idealized, fixed aggregation rather than a trace of the
deployed governance protocol: it assigns one equal-weight vote to each of
the six methods and approves a skill only when at
least two-thirds vote \texttt{safe}. To isolate each method's marginal
contribution, we run a leave-one-out ablation that removes one
auditor at a time and recomputes the verdict over the remaining five
with a 3-of-5 majority.

Figure~\ref{fig:rq3-metrics-six-panel} shows that \textbf{SkillScan}
is the strongest standalone auditor, achieving 96.0\% accuracy and
97.1\% F1-score with no false positives. Other methods exhibit
complementary weaknesses: \textbf{Static} and \textbf{Manifest}
over-reject benign skills, while \textbf{Security Scanner} misses
many malicious skills. By aggregating these complementary signals,
the DAO Committee improves the overall operating point to 97.6\%
accuracy, 98.5\% recall, 98.3\% F1-score, and a 1.5\% FN rate.
These results indicate that aggregation reduces method-specific blind
spots rather than merely matching the best standalone auditor.
This complementarity supports the pluggable DAO design of
\S\ref{sec:design:dao}.

Table~\ref{tab:rq3-auditor-ablation} reports the leave-one-out
ablation. Precision remains above 98\% (98.02\%--99.96\%), F1-score
above 95.5\%, and accuracy above 93.8\% in every variant, so no
single removal collapses the verdict. Every removal nonetheless
reduces F1 relative to full \sysname (98.32\%): removing
\textbf{SkillScan} is most damaging (93.14\% recall, 95.52\% F1,
6.86\% FN rate); removing \textbf{Static} or \textbf{Manifest}
improves precision (99.95\%--99.96\%) and FP rate (0.10\%--0.13\%)
at the cost of about 4.6\,pp recall; removing \textbf{Security
Scanner} has the smallest effect (98.17\% F1).

\begin{summarybox}
\noindent\textbf{Answer to RQ3:}\quad The evaluated fixed six-method
aggregation outperforms
every standalone method (97.6\% accuracy, 1.5\% FN rate vs.
SkillScan's 96.0\% / 5.7\%), and the leave-one-out ablation shows
every method contributes. These results do not directly estimate the
accuracy of a deployed committee in which reputation-weighted auditors
choose heterogeneous methods.
\end{summarybox}

\subsection{Incentive Compatibility}
\label{sec:eval-rq4}

An audit committee remains vulnerable if dishonest voting is more
profitable than truthful voting. \sysname's economic mechanism
(\S\ref{sec:design:economics}) must therefore make honest auditing
strictly preferable in both a single round and over many rounds.

We test this in two complementary ways. A one-shot game-theoretic
analysis identifies the equilibrium of a single audit round under
\sysname's reward and slash rules; a multi-round simulation then
tracks stake and reputation across 600 rounds under the full update
rule, including reward, slash, and decay.

\subsubsection{Game-Theoretic Analysis}
\label{sec:eval-rq4-game}

We model one audit round as a two-player game between a developer
$d$ and an auditor $a$. The developer submits either a benign (B) or
malicious (M) skill; the auditor either cooperates (C) by voting
truthfully or defects (D) by casting an incorrect verdict. Under the
honest-majority assumption of \S\ref{sec:eval-analysis}, the other
$N\!-\!1\!=\!19$ auditors vote truthfully, so the reputation-weighted
verdict matches ground truth and losing-side auditors are slashed.

We reuse the reward $R$ (Eq.~\ref{eq:reward}), slash
$S=\gamma R_{\mathit{base}}(j)$ with $\gamma=2$
(Eq.~\ref{eq:slash}), publication fee $C_{\mathit{pub}}(j)$
(Eq.~\ref{eq:pub-fee}), benign-skill utility $U_{\mathrm{legit}}$, and
bribe $B$. Since $S=2R_{\mathit{base}}(j)>R_{\mathit{base}}(j)\ge R$
and a rational briber pays $B<S$, we have $S>B>0$. The resulting payoff
matrix is shown in Table~\ref{tab:rq4-payoff}.

\begin{table}[t]
\caption{One-round payoff matrix $(U_d, U_a)$ between developer
$d$ and auditor $a$.}
\label{tab:rq4-payoff}
\centering
\small

\begin{tabular}{|c|c|c|}
\hline
\textbf{d} $\backslash$ \textbf{a} & \textbf{C} & \textbf{D} \\
\hline
\textbf{B}
  & $(U_{\mathrm{legit}}\!-\!C_{\mathit{pub}},\; +R)$
  & $(U_{\mathrm{legit}}\!-\!C_{\mathit{pub}},\; -S)$ \\
\hline
\textbf{M}
  & $(-C_{\mathit{pub}},\; +R)$
  & $(-C_{\mathit{pub}}\!-\!B,\; B\!-\!S)$ \\
\hline
\end{tabular}
\vspace{2pt}
\parbox{\linewidth}{\footnotesize $R$: consensus-aligned reward;
$S$: slash with $S>R_{\mathit{base}}(j)\ge R>B$; $C_{\mathit{pub}}$:
publication fee; $U_{\mathrm{legit}}$: developer revenue from an
approved skill; $B$: bribe to a defecting auditor.}
\end{table}

Under these parameters, truthful auditing is the only stable auditor
strategy. For benign skills, $C$ receives $+R$, whereas $D$
incurs a slash of $-S$. For malicious skills, C receives $+R$, whereas D
receives $B-S<0$. Thus, $C$ strictly dominates $D$ for both benign and malicious submissions.
For the developer, M yields at
most $-C_{\mathit{pub}}$, while B yields
$U_{\mathrm{legit}}-C_{\mathit{pub}}>0$ whenever the skill has positive
legitimate value. Therefore $(\text{B},\text{C})$ is the unique Nash
equilibrium; any unilateral deviation loses at least
$\min\{U_{\mathrm{legit}},S-B\}>0$.

\subsubsection{Multi-Round Simulation}
\label{sec:eval-rq4-sim}

The one-shot result does not capture accumulated stake and
reputation updates. We therefore simulate $N\!=\!20$ auditors (15
honest, 5 malicious) over 600 rounds, with 5 auditors sampled per
round. We set $R_{\mathit{base}}=0.56$~TC, $\gamma=2$,
$r_i=r_{\max}$, and $\varphi_{\mathit{proto}}\to0$, which gives a
$+0.56$~TC reward for correct votes and a 1.12~TC slash for
incorrect votes against an initial 100~TC balance. Reputation starts
at $r_0\!=\!100$ and uses $\delta^{+}\!=\!+15$,
$\delta^{-}\!=\!-30$, and decay $\alpha\!=\!0.995$. We compare four
auditor types: fully honest, 80\% honest, default malicious
(30\% correct votes), and stealthy malicious (defects in 50\% of
the rounds in which it is sampled).

\begin{figure}[t]
  \centering
  \includegraphics[width=0.9\linewidth]{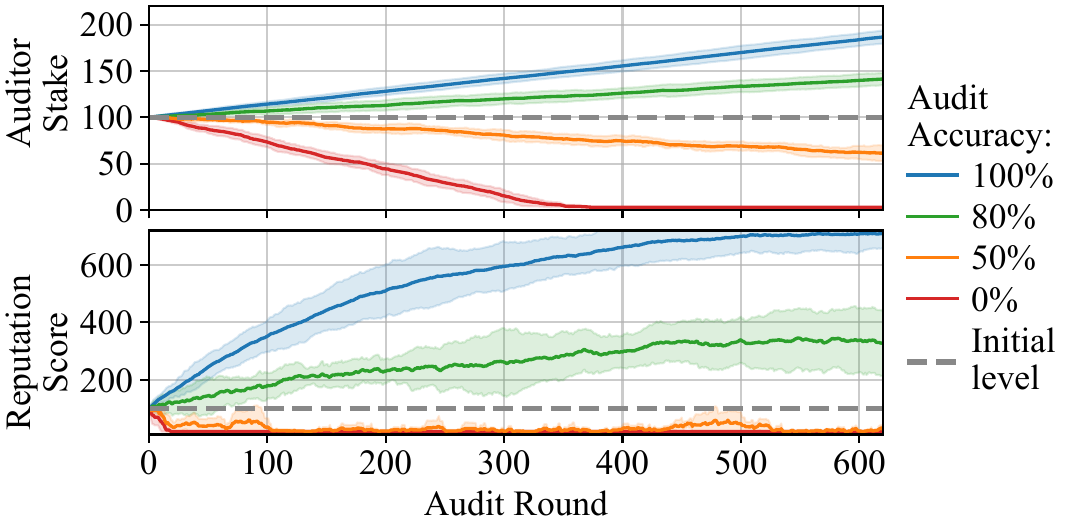}
  \caption{The TC balance (top) and reputation (bottom) over 600 audit
  rounds.}
  \Description{A two-panel convergence plot compares fully honest auditors, 80 percent correct honest auditors, default malicious auditors, and malicious auditors that misbehave in 50 percent of sampled audits. The top panel shows stake trajectories and the bottom panel shows reputation trajectories over 600 audit rounds. Honest auditors retain or increase stake and reputation, while malicious auditors lose stake and reputation over repeated audit rounds.}
  \label{fig:rq4-convergence}
\end{figure}

Figure~\ref{fig:rq4-convergence} shows a clear separation. Fully
honest auditors grow from 100~TC to 189.25~TC after 600 rounds, and
the 80\%-correct cohort remains viable at 147.66~TC. The default
malicious cohort is depleted to 0~TC, and the stealthy malicious
cohort falls to 34.59~TC. Reputation follows the same pattern:
fully honest auditors reach 744.84, the 80\%-correct cohort reaches
327.24, and the malicious cohorts collapse to 5.96 and 11.69.

The game and the simulation jointly validate the mechanism of
\S\ref{sec:design:economics}: honest auditors absorb realistic
errors, while repeated malicious voting drains both stake and
reputation. With the \textbf{auditor collusion} analysis in
\S\ref{sec:eval-rq1}, the results show that malicious auditors
cannot participate sustainably under \sysname's rules.

\begin{summarybox}
\noindent\textbf{Answer to RQ4:}\quad Honest auditing is the unique
one-shot Nash equilibrium and the dominant long-run strategy: across
600 simulated rounds, honest auditors gain stake and reputation while
malicious cohorts are depleted to near zero.
\end{summarybox}

\section{Discussion}
\label{sec:discuss}

\sysname narrows the audit--runtime gap for LLM skills along the
four design goals of \S\ref{sec:tm-goals}, but several aspects
deserve explicit discussion.

Defensive coverage is bounded by the capabilities and freshness of the audit methods plugged into the DAO Committee. Content auditing may miss behaviors that remain within declared permissions or evade current detection patterns, and previously approved skills may become risky as attacks evolve. Aggregating multiple methods reduces these blind spots (\S\ref{sec:eval-rq3}) without eliminating them; retrospective slashing (\S\ref{sec:design:economics}) provides reactive correction, while proactive re-auditing is left to governance.

Skill content delivery to auditors (Licensed, Sealed, and
Committed skills) relies on a confidentiality bond rather than
cryptographic enforcement. The bond and reputation reset of
\S\ref{sec:design:publication} penalize confirmed leaks but cannot
prevent disclosure once a skill is decrypted, and attributing a
leak to a specific auditor without content fingerprinting remains
an open question.

Several directions follow. Stronger auditor plugins, including
attention-based detectors such as
MindGuard~\cite{wang2025mindguard}, can extend semantic coverage;
per-auditor content fingerprinting would attribute confirmed
leaks to a specific identity. Deployment also requires the DAO
to reach a critical mass of auditors for timely reviews.

\section{Conclusion}
\label{sec:concl}

We presented \sysname, the first framework that seals the
audit--runtime gap for LLM agent skills, providing end-to-end
protection from skill submission to invocation. \sysname operates in
three stages. At \emph{Submission}, a developer commits a skill and
a DAO Audit Committee reviews it through reputation-weighted
consensus under stake-and-slash incentives. At \emph{Anchoring},
approved skills migrate into a tamper-evident Skill
Registry that supports four publication types: Transparent,
Licensed, Sealed, and Committed. At \emph{Invocation}, the SVL,
integrated into the AI Service Provider, queries the registry,
hashes content as needed, and binds every loaded skill to its
audited record before it enters the agent's context.

On 1{,}079 in-the-wild skills, \sysname is the only defense that
provides coverage across all six representative attacks: 97.6\% accuracy on explicit
injection and implicit poisoning, 95.1\% on rug pull, 100\% on
local tampering, 90.2\% mitigation on cross-skill interaction, and stable
accuracy under 20\%--40\% colluding auditors. The DAO Audit Committee
reaches 97.6\% accuracy, 98.3\% F1-score, and a 1.5\% FN rate; the SVL incurs mean invocation latencies of 51.7--157.6 ms across the four registry backends, while local validation adds at most 0.495 ms. A
game-theoretic analysis and a 600-round simulation show that
honest auditing is the unique Nash equilibrium and the dominant
long-run strategy. Together, these results show that LLM skills can
be submitted, audited, anchored, and loaded with stronger end-to-end
trust guarantees.

\section{Ethical Considerations}
\label{sec:ethics}

\sysname is a defensive system, but our use of real-world data warrants
explicit ethical discussion.

\paragraph{Data sources.}
The 1{,}079 skills we evaluate were collected entirely from
publicly accessible sources: ClawHub, public GitHub repositories,
and other open skill registries. The data contain
no personally identifiable information, no proprietary content,
and no human-subjects data, so no institutional review (IRB/ERB)
was required. We treat each skill's metadata and code as a
published artifact under its original license.

\paragraph{Handling of malicious skills.}
779 of the collected skills are labeled malicious, and some were
still live on production registries at collection time. For each
such skill that remained reachable to end users during our
analysis, we reported it through the platform's standard
disclosure channel before any wider use, and we did not attempt
exploitation against any third-party agent, user, or service.

\bibliographystyle{IEEEtran}
\bibliography{references}

\appendices

\section{Parameter Calibration}
\label{sec:param-experiments}

This appendix reports additional parameter-sensitivity experiments
for \sysname and justifies the parameter values chosen in
\S\ref{sec:design}. Unless otherwise stated, each experiment follows
the evaluation setup in \S\ref{sec:eval} and varies one parameter at
a time while keeping the remaining configuration fixed.

\subsection{Initial Reputation Ratio}
\label{sec:param-reputation}

\begin{figure}[H]
  \centering
  \includegraphics[width=0.96\columnwidth]{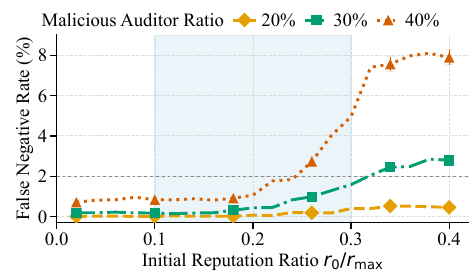}
  \caption{Sensitivity of malicious-skill FN Rate to the initial
  reputation ratio $r_0/r_{\max}$ under reputation-weighted voting.}
  \label{fig:reputation-sybil}
  \Description{A line plot shows malicious FN Rate as the initial
  reputation ratio increases from 0.02 to 0.40. Curves are shown for 20
  percent, 30 percent, and 40 percent malicious auditor ratios. FN Rate
  remain very low near 0.10 and increase at larger initial reputation
  ratios, especially under a 40 percent malicious auditor ratio.}
\end{figure}

We sweep the initial reputation ratio $r_0/r_{\max}$ around the
calibration used in \S\ref{sec:design:reputation}.
Figure~\ref{fig:reputation-sybil} reports the Sybil-resistance side of
this sweep under a five-auditor committee, approval threshold
$\theta=0.6$, and 20{,}000 Monte Carlo trials per setting. Malicious
auditors vote \texttt{safe} on a malicious skill, while honest auditors
vote \texttt{unsafe}; the plotted metric is the resulting malicious-skill
FN Rate under reputation-weighted voting.

The FN Rate remains near zero when fresh identities receive little
initial weight, and then rises as $r_0/r_{\max}$ increases. At the
calibrated value $r_0/r_{\max}=0.10$, the malicious FN Rate is only
0.18\% with 30\% malicious auditors and 0.84\% even when the malicious
auditor ratio increases to 40\%. In contrast, larger initial reputations
substantially weaken this margin: at $r_0/r_{\max}=0.30$, the FN Rate
reaches 1.58\% for 30\% malicious auditors and 4.95\% for 40\% malicious
auditors; at $r_0/r_{\max}=0.40$, it further rises to 2.79\% and 7.89\%,
respectively. Thus, the interval $r_0/r_{\max}\in[0.10,0.30]$ remains
acceptable for the 30\% malicious-auditor reference setting, but the lower
end of the interval provides a substantially larger safety margin under
stronger Sybil pressure. This supports the conservative choice
$r_0/r_{\max}=0.10$ in \S\ref{sec:design:reputation}.

\subsection{Number of Audit Methods}
\label{sec:param-audit-methods}

We next vary the number of audit methods available to the DAO Committee.
This experiment asks how many independent audit signals are needed before
the committee reaches a stable operating point. We evaluate committees with
$N=4,5,6,7$ audit methods. Because the current implementation contains six
empirical audit methods, the $N=4$ and $N=5$ cases enumerate all subsets of
the six empirical methods and report the mean across subsets; $N=6$
corresponds to the full empirical committee. The $N=7$ cases are
extrapolations that add one modeled future method on top of the six
empirical methods. The ``average plugin'' models a future method whose
per-dataset safe-vote rate matches the average behavior of existing
methods, while the ``strong plugin'' models a future method with behavior
similar to the best existing audit method. These two variants separate the
effect of adding one more audit method from the quality of the added
method. To keep the decision rule consistent with the main evaluation, a
committee with $N$ methods approves a skill when at least
$\lceil 0.6N\rceil$ methods vote \texttt{safe}. Thus, $N=4,5,6,7$
correspond to 3-of-4, 3-of-5, 4-of-6, and 5-of-7 voting rules,
respectively.

\begin{table}[t]
  \centering
  \caption{Sensitivity to the number of audit methods in the DAO Committee.
Accuracy, F1-score, FN Rate, and FP Rate are
  reported as percentages.}
  \label{tab:param-audit-methods}
  \small
  \setlength{\tabcolsep}{3.5pt}
  \renewcommand{\arraystretch}{1.12}

  \newcommand{\paramDrop}[2]{#1{\scriptsize\textcolor{red}{\,$\blacktriangledown$\,#2}}}
  \newcommand{\paramRiseBad}[2]{#1{\scriptsize\textcolor{blue}{\,$\blacktriangle$\,#2}}}
  \newcommand{\paramGain}[2]{#1{\scriptsize\textcolor{blue}{\,$\blacktriangle$\,#2}}}
  \newcommand{\paramBetterLow}[2]{#1{\scriptsize\textcolor{red}{\,$\blacktriangledown$\,#2}}}
  \rowcolors{0}{gray!12}{white}

  \resizebox{\columnwidth}{!}{%
  \begin{tabular}{@{}lcrrrr@{}}
    \toprule
    \rowcolor{white}
    \textbf{Setting} & \textbf{Rule} & \textbf{Acc.} & \textbf{F1} &
    \textbf{FN Rate} & \textbf{FP Rate} \\
    \midrule
    $N=4$ empirical subsets
    & 3-of-4
    & \paramDrop{91.96\%}{5.66\%}
    & \paramDrop{94.66\%}{3.66\%}
    & \paramRiseBad{2.37\%}{0.89\%}
    & \paramRiseBad{21.69\%}{17.17\%} \\

    $N=5$ empirical subsets
    & 3-of-5
    & \paramDrop{95.78\%}{1.84\%}
    & \paramDrop{96.95\%}{1.37\%}
    & \paramRiseBad{5.02\%}{3.54\%}
    & \paramBetterLow{2.30\%}{2.22\%} \\

    \midrule
    \rowcolor{white}
    \textbf{$N=6$ full committee}
    & \textbf{4-of-6}
    & \textbf{97.62\%}
    & \textbf{98.32\%}
    & \textbf{1.48\%}
    & \textbf{4.52\%} \\
    \midrule

    $N=7$ average plugin
    & 5-of-7
    & \paramDrop{95.06\%}{2.57\%}
    & \paramDrop{96.60\%}{1.72\%}
    & \paramBetterLow{0.59\%}{0.89\%}
    & \paramRiseBad{15.43\%}{10.91\%} \\

    $N=7$ strong plugin
    & 5-of-7
    & \paramGain{98.42\%}{0.80\%}
    & \paramGain{98.89\%}{0.57\%}
    & \paramBetterLow{0.36\%}{1.12\%}
    & 4.52\%{\scriptsize\textcolor{gray}{\,--}} \\
    \bottomrule
  \end{tabular}%
  }
  \rowcolors{0}{}{}
  \vspace{2pt}
  \parbox{\linewidth}{\footnotesize
  \textcolor{red}{$\blacktriangledown$}: decrease;
  \textcolor{blue}{$\blacktriangle$}: increase, relative to the
  $N=6$ full committee.}
\end{table}

The results show that adding an audit method is beneficial only when the
new method provides high-quality and complementary evidence. Four methods
are not sufficient as a robust default: the average FP Rate remains high
(21.69\%), indicating frequent benign-skill rejections when the committee
has limited independent evidence. Moving to five methods reduces the FP
rate to 2.30\%, but increases the FN Rate to 5.02\%, indicating a weaker
safety margin against malicious skills. The six-method committee provides
the best empirically supported trade-off, achieving the highest measured
accuracy and F1-score among real configurations while keeping FN
Rate and FP Rate low. A modeled seventh method can improve safety-oriented metrics
when it is strong, but an average-quality plugin increases the FP Rate and
reduces overall performance. Therefore, \sysname uses the six-method
committee as the default deployment setting and admits additional audit
methods only when validation shows that they improve coverage without
substantially increasing benign-skill rejection.

\subsection{Slash Coefficient}
\label{sec:param-slash}

\begin{figure}[H]
  \centering
  \includegraphics[width=0.96\columnwidth]{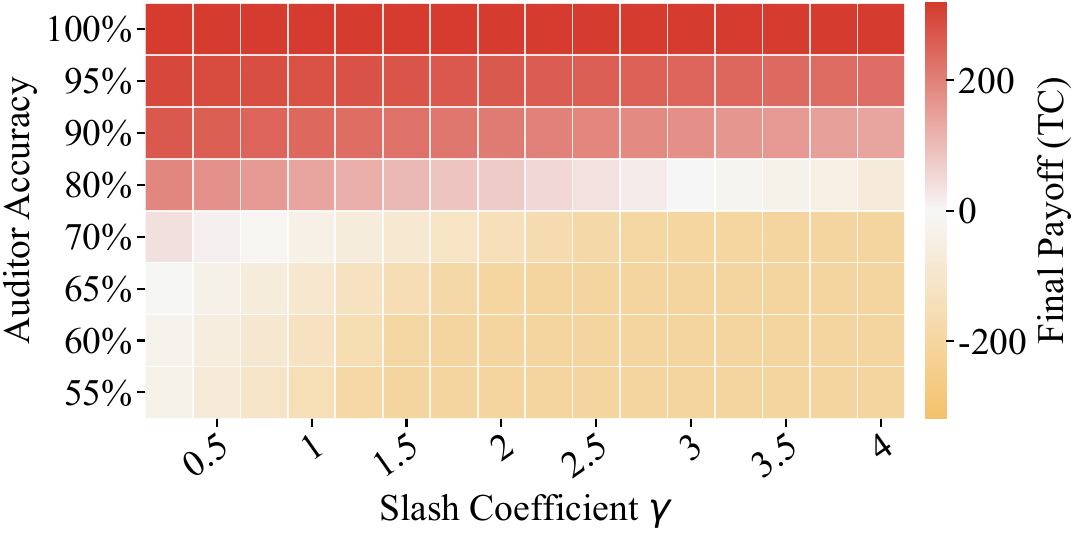}
  \caption{Sensitivity of auditor payoff to the slash coefficient $\gamma$.
  The color scale reports final payoff after 600 audit rounds for auditors
  with different correctness probabilities.}
  \label{fig:param-slash}
  \Description{A heatmap shows final auditor payoff across slash
  coefficients and auditor accuracy levels. Low-accuracy auditors have
  negative payoff across most settings, while high-accuracy auditors remain
  profitable. Larger slash coefficients reduce payoff for imperfect
  auditors.}
\end{figure}

We finally study the slash coefficient $\gamma$ in Eq.~\ref{eq:slash}.
This parameter controls the monetary penalty applied to an auditor whose
vote diverges from the final committee consensus. The value must be large
enough to make low-quality or bribed voting unattractive, but not so large
that honest auditors avoid difficult cases because occasional mistakes
create excessive downside risk. To characterize this trade-off, we sweep
$\gamma\in[0.25,4.00]$ and simulate auditors with per-round correctness
probabilities from 55\% to 100\%. Each setting runs 600 audit rounds with
$R_{\mathit{base}}=0.56$~TC, an initial balance of 200~TC, $r_0=100$,
$r_{\max}=1000$, $\delta^{+}=+15$, $\delta^{-}=-30$, and reputation decay
$\alpha=0.995$. Figure~\ref{fig:param-slash} reports the final payoff,
measured as the final TC balance minus the initial balance.

The sweep identifies a narrow useful range for $\gamma$. Small slash
coefficients provide weak deterrence: when $\gamma=1.5$, auditors with only
65\% accuracy can still remain profitable, making the system more tolerant
of unreliable or strategically biased voting. In contrast, very large
coefficients become punitive for useful but imperfect auditors; for example,
when $\gamma=4.0$, an 80\%-accuracy auditor loses 64.71~TC over 600 rounds.
The default $\gamma=2$ provides the best balance. It makes low-quality
auditors with 65\%--70\% accuracy unprofitable from cold start while still
allowing higher-quality auditors to earn positive long-run returns. Thus,
$\gamma=2$ is large enough to discourage low-quality or bribed voting while
preserving participation incentives for accurate auditors. In practice,
$\gamma\in[2,3]$ is a reasonable operating range: smaller values preserve
participation in ambiguous cases, whereas larger values impose stricter
penalties on inaccurate or strategically biased auditors.

\end{document}